\documentclass[a4paper,10pt,prd,aps,floatfix,preprintnumbers,nofootinbib,superscriptaddress]{revtex4}
\usepackage{amssymb}
\usepackage{amsmath}
\usepackage{graphicx}
\usepackage{verbatim}
\usepackage{microtype}
\usepackage{placeins}
\usepackage[dvipsnames]{xcolor}
\usepackage{orcidlink}
\usepackage{bbm}
\usepackage{braket}

\newcommand{\orcidauthorBENNETT}{0000-0002-1678-6701}
\newcommand{\orcidauthorDELDEBBIO}{0000-0003-4246-3305}
\newcommand{\orcidauthorHILL}{0000-0003-2383-940X}
\newcommand{\orcidauthorHONG}{0000-0002-3923-4184}
\newcommand{\orcidauthorVADACCHINO}{0000-0002-5783-5602}
\newcommand{\orcidauthorLEE}{0000-0002-4616-2422}
\newcommand{\orcidauthorLUCINI}{0000-0001-8974-8266}
\newcommand{\orcidauthorPIAI}{0000-0002-2251-0111} 
\newcommand{\orcidauthorPROVATAS}{0000-0002-3132-9621}
\newcommand{\orcidauthorALLTON}{0000-0003-0795-124X}
\newcommand{\orcidauthorLIN}{0000-0003-3743-0840}

\newcommand{\tint}{\tau_{\mathrm{int}}}
\newcommand{\beq}{\begin{equation}}
\newcommand{\eeq}{\end{equation}}
\newcommand{\beqs}{\begin{eqnarray}}
\newcommand{\eeqs}{\end{eqnarray}}
\newcommand{\dd}{\mbox{d}}

\newcommand{\gsim}{\mathrel{\raisebox{-
.6ex}{$\stackrel{\textstyle>}{\sim}$}}}

\makeatletter
\def\@collaboration@present#1#2#3{
 \par
 \begingroup
  \frontmatter@collaboration@above
  \@author@present{\ignorespaces#2\unskip}{#3}
  \par
 \endgroup
 \set@listcomma@list#1
}
\makeatother

\begin{document}

\preprint{ET-0835A-26}

\preprint{CPTU-PTC-26-31}

\preprint{PNUTP-26/A03}

\title{Finite temperature Sp(4) lattice gauge theory with matter fields}

\author{Chris Allton\,\orcidlink{\orcidauthorALLTON}}
\email{c.r.allton@swansea.ac.uk}
\affiliation{Department of Physics, Faculty  of Science and Engineering, Swansea University, Singleton Park, SA2 8PP, Swansea, United Kingdom}
\affiliation{Centre for Quantum Fields and Gravity, Faculty  of Science and Engineering, Swansea University, Singleton Park, SA2 8PP, Swansea, United Kingdom}

\author{Ed Bennett\,\orcidlink{\orcidauthorBENNETT}}
\email{e.j.bennett@swansea.ac.uk}
\affiliation{Centre for Quantum Fields and Gravity, Faculty  of Science and Engineering, Swansea University, Singleton Park, SA2 8PP, Swansea, United Kingdom}
\affiliation{Swansea Academy of Advanced Computing, Swansea University (Bay Campus), Fabian Way, Swansea SA1 8EN, United Kingdom}

\author{Luigi Del Debbio\,\orcidlink{\orcidauthorDELDEBBIO}}
\email{luigi.del.debbio@ed.ac.uk}
\affiliation{Higgs Centre for Theoretical Physics, School of Physics and Astronomy, The University of Edinburgh, Peter Guthrie Tait Road, Edinburgh EH9 3FD, United Kingdom}

\author{Ryan C. Hill\,\orcidlink{\orcidauthorHILL}}
\email{ryan.hill@ed.ac.uk}
\affiliation{Higgs Centre for Theoretical Physics, School of Physics and Astronomy, The University of Edinburgh, Peter Guthrie Tait Road, Edinburgh EH9 3FD, United Kingdom}

\author{Deog Ki Hong\,\orcidlink{\orcidauthorHONG}}
\email{dkhong@pusan.ac.kr}
\affiliation{Department of Physics, Pusan National University, Busan 46241, Korea}
\affiliation{Extreme Physics Institute, Pusan National University, Busan 46241, Korea}

\author{Jong-Wan Lee\,\orcidlink{\orcidauthorLEE}}
\email{j.w.lee@ibs.re.kr}
\affiliation{Particle Theory and Cosmology Group, Center for Theoretical Physics of the Universe, Institute for Basic Science (IBS), Daejeon, 34126, Korea}

\author{C.-J. David Lin\,\orcidlink{\orcidauthorLIN}}
\email{dlin@nycu.edu.tw}
\affiliation{Institute of Physics, National Yang Ming Chiao Tung University, 1001 Ta-Hsueh Road, Hsinchu, 30010, Taiwan}
\affiliation{Centre for High Energy Physics, Chung-Yuan Christian University, Chung-Li, , 32023, Taiwan}
\affiliation{Physics Division, National Centre for Theoretical Sciences, Taipei 106319, Taiwan}

\author{Biagio Lucini\,\orcidlink{\orcidauthorLUCINI}}
\email{b.lucini@qmul.ac.uk}
\affiliation{School of Mathematical Sciences, Queen Mary University of London, Mile End Road, London, E1 4NS, UK}

\author{Maurizio Piai\,\orcidlink{\orcidauthorPIAI}}
\email{m.piai@swansea.ac.uk}
\affiliation{Department of Physics, Faculty  of Science and Engineering, Swansea University, Singleton Park, SA2 8PP, Swansea, United Kingdom}
\affiliation{Centre for Quantum Fields and Gravity, Faculty  of Science and Engineering, Swansea University, Singleton Park, SA2 8PP, Swansea, United Kingdom}

\author{Davide Vadacchino\,\orcidlink{\orcidauthorVADACCHINO}}
\email{davide.vadacchino@plymouth.ac.uk}
\affiliation{Centre for Mathematical Sciences, University of Plymouth, Plymouth, PL4 8AA, United Kingdom}

\author{Alexis Verney-Provatas\,\orcidlink{\orcidauthorPROVATAS}}
\email{2414441@swansea.ac.uk}
\affiliation{Department of Physics, Faculty  of Science and Engineering, Swansea University, Singleton Park, SA2 8PP, Swansea, United Kingdom}
\affiliation{Centre for Quantum Fields and Gravity, Faculty  of Science and Engineering, Swansea University, Singleton Park, SA2 8PP, Swansea, United Kingdom}
\affiliation{Higgs Centre for Theoretical Physics, School of Physics and Astronomy, The University of Edinburgh, Peter Guthrie Tait Road, Edinburgh EH9 3FD, United Kingdom}

\collaboration{(on behalf of the TELOS collaboration) \\ 
\href{https://telos-collaboration.github.io}{ \includegraphics[height=1cm]{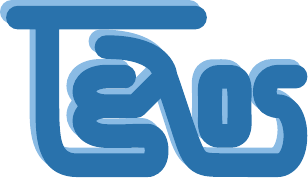} } }
\noaffiliation

\date{\today}

\begin{abstract}

We perform the first finite-temperature lattice study of the ${\rm Sp}(4)$ gauge theory coupled to two Wilson-Dirac fermions transforming in the fundamental representation. Its continuum limit plays an important role in many extensions of the Standard Model for particle physics with composite dynamics. For example, it provides a compelling realisation of the strongly interacting massive particle (SIMP) paradigm for dark matter. If realised in nature, it might lead to a first-order thermal phase transition in the early universe, and produce a detectable relic stochastic background of gravitational waves (GW). Assessing whether future experiments are sensitive to the resulting GW power spectrum requires determining the strength of the transition. We scan the parameter space of this lattice theory with a fine mesh of choices of lattice coupling, for two small values of the temporal extent of the lattice, and several choices of spatial volume and fermion mass. We measure the Polyakov loop and its susceptibility on the ensembles generated, implement reweighting, and perform a finite volume analysis, both to classify the nature of the phase transition, and  to estimate the critical coupling. We validate our algorithms and analysis procedures on the better understood ${\rm Sp}(4)$ Yang-Mills theory, by confirming the presence of a first-order phase transition. Upon coupling the theory to matter fields, in the presence of moderately light fermions, the Polyakov loop  shows a clear crossover behaviour and does not detect singularities associated with a phase transition.  We also uncover the first evidence of a first-order thermal phase transition in the case of the largest fermion masses available. This finding motivates future, large-scale studies aimed at characterising the transition in the continuum limit, for this and closely related theories with ${\rm Sp}(2N)$ gauge group. 
    
\end{abstract}

\maketitle

\tableofcontents

\section{Introduction}

Present and future generations of ground-based and space-based  experiments, designed to detect gravitational waves~\cite{Seto:2001qf,
 Kawamura:2006up,Crowder:2005nr,Corbin:2005ny,Harry:2006fi,
 Hild:2010id,Yagi:2011wg,Sathyaprakash:2012jk,Thrane:2013oya,
 Caprini:2015zlo,
 LISA:2017pwj,
 LIGOScientific:2016wof,Isoyama:2018rjb,Baker:2019nia,
 Brdar:2018num,Reitze:2019iox,Caprini:2019egz}, such as the  Einstein Telescope~\cite{Maggiore:2019uih}, offer unprecedented opportunities for step changes in our understanding of fundamental physics. They can detect the relic stochastic background of gravitational waves emitted in association with first-order phase transitions taking place in the primordial stages of evolution of our universe~\cite{Witten:1984rs,Kamionkowski:1993fg, Allen:1996vm,Schwaller:2015tja, Croon:2018erz, Christensen:2018iqi}. Prominently, many strongly coupled, asymptotically free, non-Abelian gauge theories coupled to matter fields undergo a thermal phase transition at high temperature. Some of them provide the foundations for the Strongly Interacting Massive Particle (SIMP) scenario~\cite{Hochberg:2014dra,Hochberg:2014kqa,Hochberg:2015vrg,Bernal:2017mqb,Berlin:2018tvf,Bernal:2019uqr,Tsai:2020vpi,Kondo:2022lgg,Chu:2024rrv,Samart:2025biw} for the origin of composite dark matter models, that can explain the compelling  observational evidence of dark matter (see the review in Ref.~\cite{Cirelli:2024ssz}, and references therein). In particular, they can address small-scale puzzles, such as the {\it `core vs. cusp'}~\cite{deBlok:2009sp} and the {\it `too big to fail'}~\cite{Boylan-Kolchin:2011qkt} problems. Whether these new theories are testable via gravitational wave detection depends on the detailed properties of the phase transition: its scale, order, strength, and duration, are all necessary input in software tools (such as PTPlot~\cite{Caprini:2019egz}), that compute numerically the power spectrum under some simplifying assumptions. More generally, the characterisation of the transition is necessary to determine the out-of-equilibrium dynamics responsible for bubble nucleation and evolution (see e.g., the recent Refs.~\cite{Blasi:2026lld,Rummukainen:2026etj,Rindlisbacher:2025dqw,Correia:2025qif,Huber:2025qbl} and references therein). A systematic programme aimed at assessing the experimental reach of this approach is outlined in Refs.~\cite{Huang:2020crf, Halverson:2020xpg, Kang:2021epo, Reichert:2021cvs, Reichert:2022naa, Pasechnik:2023hwv}---although estimating directly the nucleation rate remains challenging~\cite{Kalikmanov2013}.

The ${\rm Sp}(4)$ gauge theory coupled to $N_f=2$ species of Dirac fermions transforming in the fundamental representation is the prototype for an ultraviolet completion of SIMP models~\cite{Hochberg:2014kqa,Hochberg:2018rjs,Tsai:2020vpi,Maas:2021gbf,Zierler:2021cfa,Kulkarni:2022bvh,Pomper:2024otb}. The theory is special because of its enhanced global ${\rm SU}(4)$ symmetry, that  acts on the matter fields, and that is broken to its maximal ${\rm Sp}(4)$ subgroup in the presence of degenerate fermion masses. As the formation of condensates breaks the global symmetry also spontaneously, the lightest bound states are the Pseudo-Nambu-Goldstone Bosons (PNGBs) spanning the ${\rm SU}(4)/{\rm Sp}(4)$ coset. They act as dark matter candidates, and the relic density is governed by $3\rightarrow 2$ processes between PNGBs. Phenomenologically,  explaining the dark matter distribution in galaxies~\cite{Kaplinghat:2015aga}, and the early universe evolution of dark matter~\cite{Bernreuther:2023kcg,Chu:2024rrv,Samart:2025biw}, requires precise knowledge of the masses and couplings of the bound states of the theory, including resonances~\cite{Berlin:2018tvf}. This requirement stimulated a number of exploratory spectroscopy studies using lattice field theory~\cite{Bennett:2023rsl,Dengler:2024maq,TELOS:2026mim,TELOS:2026alk}. The natural complement for the measurement of masses and couplings at zero temperature is to study the behaviour of the theory at high temperatures, that mimic the conditions in the early universe, and characterise its thermal phase transition, in view of the  aforementioned GW experimental programme.

The study of finite temperature behaviour and phase structure of the field theories relevant to particle physics is a rich field. The story of how it reached its present stage of maturity is interesting not just in itself, but also for the technological developments it stimulated. A first-order phase transition in the early universe would fulfill Sakharov's out-of-equilibrium condition~\cite{Sakharov:1967dj}, a necessary requirement for baryogenesis to provide an explanation of the present day matter-antimatter asymmetry of the universe. The failure of the Standard Model (SM) of particle physics to meet this requirement~\cite{Kajantie:1996mn, Laine:2012jy} (see  also Refs.~\cite{Laine:1998jb,Morrissey:2012db,Gould:2022ran} and references therein) is one of the compelling phenomenological arguments, together with evidence for dark matter,  demonstrating the incompleteness of the SM description of fundamental interactions. To extract reliable information about the Standard Model, leading to such a far reaching conclusion, required overcoming huge technical challenges, both  field theoretical as well as algorithmic in nature. This is particularly true for Quantum  Chromodynamics (QCD), given its strongly coupled dynamics, for which the lattice is the natural choice of research tool.

The collective effort of the lattice community has led to impressive achievements in the understanding of  finite-temperature QCD and closely related theories. The current status of lattice studies of the ${\rm SU}(3)$ Yang-Mills theory, which is known to undergo a first-order deconfinement phase transition~\cite{Svetitsky:1982gs,Yaffe:1982qf}, is critically discussed in Ref.~\cite{Borsanyi:2022xml}.  A selection of publications on the thermodynamics of ${\rm SU}(3)$ Yang-Mills theories includes also Refs.~\cite{Kajantie:1981wh,Celik:1983wz,Kogut:1983mn, Svetitsky:1983bq,Gottlieb:1985ug,Brown:1988qe,Fukugita:1989yb,Bacilieri:1989ir, Alves:1990pn, Boyd:1995zg,Boyd:1996bx, Borsanyi:2012ve,Shirogane:2016zbf} and references therein. The addition of heavy quarks, which according to the Columbia plot~\cite{Brown:1990ev,Cuteri:2021ikv} introduces an intriguing structure, has been investigated with increasing precision~\cite{Alexandrou:1998wv,Saito:2011fs,Ejiri:2019csa,Kiyohara:2021smr, Fromm:2011qi,Cuteri:2020yke,Kara:2021btt}. The status of finite temperature QCD is discussed in the review in Ref.~\cite{Aarts:2023vsf}, which provides a broad description of current trends and future avenues of research, as well as a wealth of relevant references.

The algorithmic and software developments put in place for this undertaking have been deployed to study other finite-temperature gauge theories. The pure ${\rm SU}(2)$ Yang-Mills theory is suitable for testing new machine-learning algorithms~\cite{Wetzel:2017ooo,Boyda:2020nfh}. The ${\rm SU}(2)$ theory coupled to fermions is interesting, because among other things, it avoids the sign problem, and because of its enlarged symmetries, as discussed for instance in Refs.~\cite{Hands:2006ve,Hands:2006ec,Cotter:2012mb,Lee:2017uvl, Iida:2020emi,Hands:2012fs}.  The ${\rm SU}(N_c)$ theories are interesting for their connection to gauge-gravity dualities, through the large-$N_c$ expansion. Their finite-temperature behaviour has been the subject of extensive studies~\cite{Lucini:2002ku,Lucini:2003zr,Lucini:2005vg,Panero:2009tv,Datta:2010sq,Lucini:2012wq}. Field and group theory properties, for instance the relation between the centre of the gauge group and confinement, have motivated the study of Yang-Mills theories with ${\rm G}_2$ group~\cite{Pepe:2005sz,Pepe:2006er,Cossu:2007dk,Bruno:2014rxa}, and those with ${\rm Sp}(2N)$ group~\cite{Holland:2003kg,Bruno:2024dha}. In the context of phenomenologically motivated extensions of the Standard Model, the LSD collaboration has been developing a substantial body of work on the  ${\rm SU}(4)$ theories coupled to matter fields~\cite{LatticeStrongDynamics:2020jwi,Ayyar:2024dmt,Ayyar:2025eim,Ayyar:2026wht}, as well as providing preliminary results for the ${\rm SU}(3)$ theories with large number of fermions, $N_{\rm f}$~\cite{Schaich:2015psa}---see also the review in Ref.~\cite{Asadi:2026mip}.  The challenges intrinsic to the study of this diverse set of lattice theories have also stimulated, in turn, the development and testing of alternative new algorithms, which can be applied to theories that are more closely related to QCD. One prominent example is the Logarithmic  Linear Relaxation (LLR) method~\cite{Langfeld:2012ah, Langfeld:2013xbf,Langfeld:2015fua,Cossu:2021bgn}---see for instance the lattice studies in Refs.~\cite{Springer:2021liy,Mason:2022trc,Mason:2022aka,Springer:2023wok, Lucini:2023irm,Mason:2023ixv,Bennett:2024bhy,Mason:2024dve,Bennett:2025whm,Zierler:2026hyt}. 

No lattice studies exist to date of the finite temperature behaviour of the ${\rm Sp}(4)$,  $N_{\rm f}=2$ theory. While its global symmetry breaking pattern coincides with that of the ${\rm SU}(2)$, $N_{\rm f}=2$ theory, the phase structures differ, as the corresponding Yang-Mills theories undergo a phase transition that is of second order in ${\rm SU}(2)$, but of first order in ${\rm Sp}(4)$. In particular, in the presence of large masses for the fermions one seeks evidence of a first-order phase transition, as in the Yang-Mills limit, that at lower masses turns into a smooth crossover. As this is a compelling SIMP dark-matter model, demonstrating this pattern, and characterising the strength of the transition, is propaedeutic to assessing whether it might be testable by the Einstein Telescope or other GW experiments. 

The TELOS Collaboration has recently been carrying out an extensive programme of lattice studies of the ${\rm Sp}(2N)$ theories, with and without coupling to matter fields~\cite{Bennett:2017kga,Lee:2018ztv, Bennett:2019jzz, Bennett:2019cxd, Bennett:2020hqd, Bennett:2020qtj, Bennett:2022yfa, Bennett:2022gdz, Bennett:2022ftz, Bennett:2023wjw, Bennett:2023gbe, Bennett:2023mhh, Bennett:2023qwx, Bennett:2024cqv, Bennett:2024wda, Bennett:2024tex,  TELOS:2025ash, Bennett:2025whm, TELOS:2026mim,TELOS:2026alk, Bennett:2026cev,Simonetti:2026jnx,Bennett:2026singlet}---see also Refs.~\cite{Hong:2017suj,Kulkarni:2022bvh,Bennett:2023rsl, Dengler:2024maq, Bennett:2024bhy,Lucini:2026aot}. Besides their potential to explain the origins of dark matter, these theories provide other striking opportunities for extensions of the Standard Model with composite dynamics. For example they are relevant to composite Higgs Models (CHMs)~\cite{Kaplan:1983fs,Georgi:1984af,Dugan:1984hq}---see also Refs.~\cite{Barnard:2013zea,Ferretti:2013kya,Ferretti:2016upr,Cacciapaglia:2019bqz} and the reviews in Refs.~\cite{Panico:2015jxa,Witzel:2019jbe,Cacciapaglia:2020kgq}---and models implementing the top partial compositeness (TPC) paradigm~\cite{Kaplan:1991dc}---see also Refs.~\cite{Grossman:1999ra,Gherghetta:2000qt,Chacko:2012sy}. 

In this paper, we present the results of our first steps towards extending to finite temperature our programme of studies of ${\rm Sp}(2N)$ theories coupled to matter fields. We adapt existing algorithms and software developed by the TELOS collaboration for ${\rm Sp}(4)$ gauge theories, and supplement them with technology borrowed from the literature on finite-temperature QCD. We adopt the Grid software environment~\cite{Boyle:2015tjk,Boyle:2016lbp,Yamaguchi:2022feu,Grid-repo}, supplemented by the TELOS implementation of the ${\rm Sp}(4)$ group~\cite{Bennett:2023gbe} for ensemble generation. We consider finite lattices, with two comparatively small values of the number of sites in the time direction, $N_t=4$ and $N_t=6$. We generate ensembles with dynamical Wilson-Dirac fermions, using the Hybrid Monte Carlo (HMC) algorithm~\cite{Duane:1987de} and scan the space of lattice parameters: the bare mass, $a m_0$, and the coupling, $\beta$. We perform an extensive study of the volume dependence of the Polyakov loop and its susceptibility, as a pseudo-order parameter of the transition, aimed at measuring the critical coupling. We implement reweighting~\cite{Ferrenberg:1988yz,Ferrenberg:1989ui}, as a tool to optimise the analysis of the susceptibility and characterise the transition. We validate our implementation of the HMC algorithm and reweighting, by applying them to the pure gauge ${\rm Sp}(4)$ Yang-Mills theory, which is known to undergo a first-order transition~\cite{Holland:2003kg}.\footnote{We have also compared our results to those obtained with other algorithms, specifically the heatbath (HB)~\cite{Creutz:1980zw,Cabibbo:1982zn,Kennedy:1985nu} supplemented by overrelaxation~\cite{Whitmer:1984he,Adler:1981sn}, and LLR ones, for the ${\rm Sp}(4)$ Yang-Mills gauge theory,  implemented with the  HiRep code~\cite{DelDebbio:2008zf,HiRepSUN,pica_github} extended to treat ${\rm Sp}(2N)$ gauge groups~\cite{HiRepSpN}.} We perform calibration and scale setting exercises, by computing the spectroscopy  of the lightest mesons (with the use of the Hadrons analysis framework~\cite{antonin_portelli_2023_8023716,Hadrons-repo}), as well as the Wilson flow and the topological charge, in ensembles representative of the zero-temperature behaviour of the system. We apply reweighting to analyse in detail the behaviour of the theory for lattices with $N_t=4$. Our broad scan of parameter space in the $N_t=6$ case represents a first step toward the continuum limit. We anticipate here our main result: we present the first numerical evidence that a first-order transition persists  in ${\rm Sp}(4)$ gauge theories in the presence of heavy fermions, while a crossover appears for light masses.

The paper is organised as follows.  In Sect.~\ref{Sec:continuum} we recall the salient properties of the continuum theory of interest. Our choice of lattice action, the algorithms we deploy to study it, and the observables of interest are discussed in Sect.~\ref{Sec:lattice}. We summarise and critically discuss our numerical results in Sect.~\ref{Sec:Results}. We outline a systematic programme of study of this and related theories in Sect.~\ref{Sec:outlook}. Some technical details that are useful for reproducibility purposes, and pertain to our study of autocorrelation as a way to determine the optimal size of our ensembles, are relegated to Appendix~\ref{App:Autocorrelation}. Additional tests of the transition, conducted with the Binder cumulant, are reported in Appendix~\ref{Sec:Binder}.

\section{Quantum field theory}
\label{Sec:continuum}

In this section, we describe the continuum theory of interest, by explicitly defining its Lagrangian density and discussing its salient global symmetry properties. We then briefly introduce its finite temperature formulation. To provide context, we summarise also what is known about the finite-temperature thermal transition in other, closely related, gauge theories coupled to matter fields transforming in the fundamental representation, as well as about Yang-Mills theories with other gauge groups, but no matter fields.

\subsection{Lagrangian density and symmetries}
\label{Sec:Lagrangian}

To define the symplectic group, ${\rm Sp}(2N)$, we introduce the  $2N\times2N$ (skew-antisymmetric) symplectic tensor, $\Omega$, which we write in the following block-matrix form:
\begin{equation}
    \Omega = \left(
    \begin{array}{c|c}
        \mathbb{O}_{N\times N} & \mathbbm{1}_{N\times N} \\
        \hline
        - \mathbbm{1}_{N\times N} & \mathbb{O}_{N\times N}
    \end{array}\right)\,,
    \label{eq:DefinitionAntisymmetricMatrix}
\end{equation}
The elements of ${\rm Sp}(2N)$ are those ${\rm SU}(2N)$ matrices that satisfy the following commutation relation with $\Omega$:
\begin{equation}
    U \Omega U ^T= \Omega\,.
    \label{eq:DefinitionSp2N}
\end{equation}
The  $N(2N+1)$ generators of the group are denoted as $t^A$, in which the index, $A=1,\,\cdots,\, N(2N+1)$, labels the adjoint representation of ${\rm Sp}(2N)$. We choose to work with a basis in which the Hermitian and symmetric (real) generators are normalised so that $\mathrm{Tr}\left[t^At^B\right]=\frac{1}{2} \delta^{AB}$.

The theory of interest in this paper has gauge group ${\rm Sp}(2N=4)$ and includes $N_f=2$ Dirac fermions, that transform in the fundamental representation of the gauge group, and have degenerate mass, $m^{\rm f}$. The continuum Lagrangian density in Minkowski space can be written as follows
\begin{equation}
\begin{aligned}
    \mathcal{L} = -\frac{1}{2} &\mathrm{Tr} \left[ G_{\mu\nu}\frac{}{}G^{\mu\nu}\right]
    - m^{\rm f} \sum_{J=1}^{N_f} \left(\overline{Q^J}\right)_a Q^{J\, a} \\
    + \frac{1}{2} & \sum_{J=1}^{N_f} \left[
    i \left(\overline{Q^{J}}\right)_a \gamma^\mu\left( D_\mu Q^{J}\right)^{a}
    - i \left(\overline{D_\mu Q^{J}}\right)_a \gamma^\mu Q^{J\, a}
   \right]\,,
    \label{eq:ContinuumLagrangian}
\end{aligned}
\end{equation}
where the bar on the spinor denotes the Dirac adjoint,
$\overline Q\equiv Q^{\dagger}\gamma^0$. In this expression we use Greek letters, $\mu,\,\nu$, to  denote Lorentz indices,  Roman letters, $a$, to denote the index in the fundamental representation of the gauge group, and capitalised Roman letters, $J$,  to denote the flavour  index. We suppress spinor indices, which are summed over. The gauge covariant derivative, $D_\mu Q^J$, and field-strength tensor, $G_{\mu\nu}$,  are defined as follows:
\begin{align}
    D_\mu Q^J &\equiv \partial_\mu Q^J\,+\, i g A_\mu Q^J\,,
    \\
    G_{\mu\nu} &\equiv \partial_\mu A_\nu - \partial_\nu A_\mu + i g [A_\mu, A_\nu]\,,
\end{align}
where $g$ is the gauge coupling, the gauge fields are written in terms of the generators of the ${\rm Sp}(4)$ group, as the $4\times 4$ matrices  $A_{\mu}\equiv \sum_AA_{\mu}^At^A$, and $\left[\cdot\,,\,\cdot\right]$ is the commutator between matrices.

\begin{table}[t]
    \centering
    \caption{Flavoured meson operators, $\mathcal{O}_{\rm M}$, used in the study of the ${\rm Sp}(4)$ gauge theory coupled to  $N_{\rm f}=2$ Dirac fermions, $Q^J$, transforming in the fundamental representation. Gauge and spinor indices are understood. We report the spin and parity quantum numbers, $J^P$, of the operators and particles they source, as well as their transformation properties under the global, unbroken ${\rm Sp}(4)$ group, following the conventions in Ref.~\cite{Bennett:2024cqv}.\\}
    \begin{tabular}{ |c|c|c|c|c| }
        \hline\hline
        $~~~~$Label,  ${\rm M}~~~~$ &   $~~~~$Operator, $\mathcal{O}_{\rm M}$  $~~~~$ & $~~~~J^P~~~~$ & $~~~~Sp(4)~~~~$  \\
        \hline
        PS & $ \overline{Q^1} \gamma_5 Q^2$ & $0^{-}$ & $\textbf{5}$  \\
        V  & $ \overline{Q^1} \gamma_\mu Q^2$ & $1^{-}$ & $\textbf{10}$  \\
        \hline\hline
    \end{tabular}
    \label{tab:mesons}
\end{table}

Our interest in the gauge theories with ${\rm Sp}(2N)$ groups is motivated by the pseudoreality of the fundamental representation, which leads to enhanced global symmetries acting on the fermions. The global ${\rm U}(2)_L \times {\rm U}(2)_R$  symmetry, acting on the left-handed and right-handed projections of the fermions, is enhanced to a ${\rm U}(1)_A\times {\rm SU}(4)$ symmetry. The $U(1)_A$ is anomalous, and we can ignore it for the purposes of this paper.  The ${\rm SU}(4)$ group is broken to its ${\rm Sp}(4)$ subgroup, both explicitly, in the presence of non-vanishing mass, $m^{\rm f}\neq 0$, and spontaneously, by the dynamical formation of the associated fermion condensate. As a result, five PNGBs form, as light composite states, transforming along the directions of the coset, ${\rm SU}(4)/{\rm Sp}(4)$. In general, all the bound states of the theory organise themselves into degenerate multiplets of the unbroken, global ${\rm Sp}(4)$ symmetry group. The examples discussed later in the paper are the pseudoscalar (${\rm PS}$) and vector (${\rm V}$) mesons, transforming, respectively, as the $\textbf{5}$ and $\textbf{10}$ irreducible representations---see Table~\ref{tab:mesons}, in which we report also the explicit form of the associated interpolating operators, ${\cal O}_{\rm M}$, as well as the spin and parity of the particles they source, $J^P$.

This theory plays a central role in the context of composite Higgs models (CHM)~\cite{Barnard:2013zea,Ferretti:2013kya,Ferretti:2016upr,Cacciapaglia:2019bqz}, in which the Higgs fields emerge as part of the multiplet of PNGBs from the underlying dynamics. The same ${\rm Sp}(4)$ theory with $N_f=2$ fundamental Dirac fermions is prominent also in the context of SIMP dark matter~\cite{Hochberg:2014dra,Hochberg:2014kqa,Hochberg:2015vrg}, because in relevant portions of parameter space it allows for the presence of $3\rightarrow 2$ processes that affect the phenomenology. The TELOS collaboration has studied the spectrum and other zero-temperature properties of the bound states in this theory, both in the quenched limit~\cite{Bennett:2017kga,Bennett:2019cxd,Bennett:2023qwx}, as in the presence of dynamical Wilson~\cite{Bennett:2019jzz,Bennett:2023rsl,TELOS:2026alk} and M\"{o}bius Domain Wall Fermions (MDWFs)~\cite{Bennett:2026cev}. For this exploratory study, we employ Wilson-Dirac fermions, as described in Sect.~\ref{Sec:lattice}.

\subsection{Elements of finite temperature field theory}
\label{Sec:finiteT}

We briefly recall  the basic principles underlying the study of the equilibrium properties of  quantum field theories in contact with a heat bath at finite temperature, $T$, and the study of their phase transitions. More extensive and pedagogical discussions can be found in textbooks (see, e.g., Refs.~\cite{Rothe:1992nt,Gattringer:2010zz}) and reviews (see, e.g., Refs.~\cite{Philipsen:2021qji,Guenther:2020jwe,Aarts:2023vsf}).

Given a quantum field theory, living in four dimensions, governed by the classical action obtained by integrating over the (Minkowski) spacetime a Lagrangian density, ${\cal L}$, that, as the one in Eq.~(\ref{eq:ContinuumLagrangian}), depends on the fields of the theory, which we collectively denote as $\phi$, and their derivatives,   its formulation as a finite-temperature equilibrium system is obtained by applying the following prescription:
\begin{itemize}
\item perform a Wick rotation, by replacing the spacetime temporal variable, $t$, with its Euclidean counterpart, $\tau=i t$,
\item assume that $\tau$ extends over a circle with finite circumference, $\boldsymbol{\beta}=1/T$,\footnote{We use the  notation $\boldsymbol{\beta}=1/(kT)$, reminiscent of statistical mechanics, to highlight the distinction with the lattice coupling, $\beta$. Furthermore, we set the Boltzmann constant to natural units, $k=1$.} so that $0\leq \tau < 1/T$,
\item along this compact temporal direction, impose periodic boundary conditions on bosons, antiperiodic  on fermions.
\end{itemize}
If the system is bound to live inside a three-dimensional container with finite volume, $V$, then the path integral becomes a function of $T$ and $V$, and one can write it as the partition function of a statistical mechanics system:
\beqs
{\cal Z}(T,V) &\equiv & \int {\cal D} \phi \exp\left[{i\,\int \dd t \dd^3 x \,{\cal L}}\right]\,=\, 
\int {\cal D} \phi \exp\left[-\,{\int_0^{1/T} \dd \tau \int_V \dd^3 x \,{\cal L}_E}\right]
\,\equiv\,\int {\cal D} \phi \,e^{-\,{\cal S}_E}\,,
\eeqs
where ${\cal D} \phi $ stands for the path-integral measure with proper boundary conditions,  ${\cal L}_E$ is the Euclidean Lagrangian density, and ${\cal S}_E$ the Euclidean action, obtained by integrating ${\cal L}_E$ in four Euclidean dimensions.\footnote{We conventionally define the Euclidean Lagrangian density, ${\cal L}_E$, as minus the Lagrangian density, ${\cal L}$, evaluated by replacing the spacetime $t$ with the imaginary $t=-i \tau$. In the limiting case in which the Lagrangian density reduces to a constant effective potential, ${\cal L}=-{\cal V}$, then ${\cal L}_E={\cal V}$, the integrations are trivial, and one finds that, aside from a (divergent) additive constant, 
$\log \left({\cal Z}\right)\,=\,-\frac{V}{T}{\cal V}$.} Similarly, in statistical mechanics the partition function is defined by the Hamiltonian governing the microscopic dynamics, $H$, as 
\begin{equation}
  {\cal Z}(T,V)={\rm Tr}\,e^{-\boldsymbol{\beta} H}\,.  
\end{equation}
Observable quantities are evaluated as ensemble averages, given by 
\beqs
\left\langle {\cal O} \right\rangle & =&\frac{1}{\cal Z} {\rm Tr}\, \left({\cal O}\,e^{-\boldsymbol{\beta} H}\right)\equiv\frac{1}{\cal Z}\int {\cal D} \phi\,{\cal O} \,e^{-\,{\cal S}_E}\,,
\label{Eq:generalObservable}
\eeqs
for any operator ${\cal O}$. Thermodynamics is derived from the relation
\beqs
\log \left[{\cal Z}(T,V)\right]\,=\,-\frac{V}{T}{\cal F}\,,
\eeqs
where ${\cal F}=F/V$ is the (dimension-4) free energy density.\footnote{The addition to the action of a source for the time component of a conserved current is interpreted as the presence of a chemical potential, $\mu$.  In this case, the Helmholtz free energy, $F(V,\,T)$,  is replaced by the Landau free energy, $\Omega (V, \,T,\,\mu)$, and the resulting system is suitable to study the phases of hot and dense matter.}    Other thermodynamic quantities are computed as ensemble averages, that reduce to appropriate derivatives in respect to the control parameters. For example, one finds that 
\beqs
\varepsilon&\equiv& \frac{
\left\langle H \right\rangle}{V} \,=\, \frac{1}{V{\cal Z}(T,V)}\int {\cal D} \phi\,H \,e^{-\,{\cal S}_E}\,
=\,-\frac{1}{V}\frac{\partial}{\partial \boldsymbol{\beta}}\log\left[{\cal Z}(T,V)\right]\,=\,\frac{T^2}{V}\frac{\partial}{\partial T}\log\left[{\cal Z}(T,V)\right]\,,
\eeqs
which yields the internal energy density of the system, $\varepsilon$. Along the same lines, one finds that 
\beqs
P&=&-\frac{\partial F}{\partial V}\,=\,T\frac{\partial }{\partial V}\log\left[{\cal Z}(T,V)\right]\,,\quad \quad
S\,=\,-\frac{\partial F}{\partial T}\,=\,\frac{\partial }{\partial T}\left\{ T \frac{}{}\log\left[{\cal Z}(T,V)\right]\right\}\,,
\eeqs
are the pressure,  $P$, and the entropy,  $S$, respectively. For homogeneous systems, in which there are no long-distance forces and the free energy density scales trivially with the size of the system, the first such relation reduces to $P=-{\cal F}$.

\subsubsection{Deconfinement in non-Abelian gauge theories}

Broadly speaking, non-Abelian, asymptotically free, four-dimensional gauge theories are believed to exhibit confinement when probed at low energies and zero temperature. Yet, they are expected to deconfine at high temperatures. Characterising the nature and properties of this change of behaviour, in particular whether a genuine phase transition (of first or second order)  is taking place, rather than a smooth crossover, and identifying critical points in the parameter space of the theory, is an open quest. Pursuing it presents both conceptual and practical difficulties,  that descend from the non-perturbative nature of gauge theories. Lattice field theory is the instrument of choice for such a challenge.

In the case of Yang-Mills theories, in which the non-Abelian gauge fields are not coupled to matter fields, one  identifies the order parameter of the system with the Polyakov loop. This is a non-local operator, defined as
\begin{equation}
    \label{eq:ContinuumPolyakovLoop}
    L(\vec x) = \frac{1}{d_f} {\rm Tr}\left\{ {\cal P} \ \exp\left[ 
    i \int_0^{1/T}\mathrm d\tau \ A_0(\tau, \vec x)
    \right]\right\}\,,
\end{equation}
where $A_0$ is the $d_f\times d_f$, matrix-valued component of the gauge field acting on the fundamental representation (having $d_f$ components) in the temporal direction, and ${\cal P}$ is the temporal path-ordering operator. The finite-temperature, field-theory ensemble average of the absolute value of the spatially averaged Polyakov loop can be written as 
\begin{equation}
\left\langle |\frac{}{} L \frac{}{} |\right\rangle= \exp \left[  -\frac{1}{T} F_Q   \right] \,,
\end{equation} 
where $F_Q$ measures the free-energy cost of inserting an isolated, infinitely heavy static source in the fundamental representation as a function of the temperature, $T$.
The ensemble average vanishes in the confined phase, but takes non-zero values in the deconfined phase. 

This statement can be understood in terms of the centre symmetry.  The centre of the gauge group, ${\cal C}_G$, is defined by the elements of the group, $z\in G$, that commute with all the elements, $U\in G$, of the group, $z U = Uz$. In ${\rm SU}(N_c)$ groups, the centre is $\mathbb{Z}_{N_c}$, while for ${\rm Sp}(2N)$ groups it is $\mathbb{Z}_{2}$. 
The centre transformation which changes all temporal links in a timeslice, 
$U \rightarrow zU$,
leaves the (pure gauge) action invariant, however it changes the (temporal) Polyakov loop,
$ L(\vec x) \rightarrow z L(\vec x)$.
Therefore, in
Yang-Mills theories, the Polyakov loop is a good order parameter for centre symmetry breaking~\cite{Svetitsky:1982gs,Yaffe:1982qf}---see also Refs.~\cite{Holland:2000uj,Greensite:2003bk}.

It is generally accepted that four-dimensional Yang–Mills theories, with the notable exception of ${\rm SU}(2)$ gauge theory, undergo a first-order transition from the confined to the deconfined phase as the temperature increases
---see Refs.~\cite{Kajantie:1981wh,Celik:1983wz,Celik:1983nc,Kogut:1983mn, Svetitsky:1983bq,Gottlieb:1985ug,Brown:1988qe,Fukugita:1989yb, Fukugita:1989tk,Bacilieri:1989ir, Alves:1990pn, Boyd:1995zg,Boyd:1996bx, Borsanyi:2012ve,Shirogane:2016zbf,Borsanyi:2022xml,Lucini:2023irm,Giusti:2025fxu} for ${\rm SU}(3)$, Refs.~\cite{Lucini:2002ku,Lucini:2003zr,Lucini:2005vg,Panero:2009tv,Datta:2010sq,Lucini:2012wq} for ${\rm SU}(N_c)$ with $N_c>3$, Refs.~\cite{Pepe:2005sz,Pepe:2006er,Cossu:2007dk,Bruno:2014rxa} for $G_2$, and Refs.~\cite{Holland:2003kg,Bennett:2024bhy,Bruno:2024dha,Bennett:2025whm} for ${\rm Sp}(2N)$. 
For ${\rm SU}(2)$, the transition is known to be of second order~\cite{Creutz:1980zw,Engels:1989fz, Engels:1992fs, Velytsky:2007gj,Wetzel:2017ooo,Boyda:2020nfh}. 
Numerical evidence indicates that the deconfinement transition in ${\rm SU}(N_c)$ Yang–Mills theories becomes stronger with increasing $N_c$, with both the latent heat and the interface tension between the two phases, expressed in appropriate units of the critical temperature, scaling as $N_c^2$~\cite{Lucini:2005vg,Rindlisbacher:2025dqw}.

When the theory is coupled to matter fields (fermions) transforming in the fundamental representation, the centre symmetry is broken, and hence the Polyakov loop is no longer an exact order parameter. In the case of ${\rm SU}(2)$ this is believed to turn the transition into a smooth crossover, for all (finite) values of the mass---see, e.g., Refs.~\cite{Iida:2020emi, Lee:2017uvl} and references therein. Interest in this group stems from the pseudoreality of the fundamental representation of  ${\rm SU}(2)$, a property shared with the other members of the ${\rm Sp}(2N)$ family.  It implies that there is no sign problem when sampling the theory at finite density, making it amenable to numerical studies with finite chemical potential~\cite{Hands:2006ve,Hands:2006ec,Cotter:2012mb,Hands:2012fs}.

For larger gauge groups, the effect of fundamental matter fields is less straightforward. It has been verified by lattice studies that in the $SU(3)$ gauge theory with  $N_{\rm f}=2+1$ (i.e., the approximation of QCD with two degenerate light quarks and one heavier fermion representing the strange quark), and  for fermion masses close to the physical ones, the thermal transition is a smooth crossover~\cite{Aoki:2006we} (see also the more recent Refs.~\cite{Borsanyi:2010bp,Borsanyi:2010cj,HotQCD:2014kol,Aarts:2015mma,Aarts:2020vyb} and references therein). This is depicted in the famous Columbia plot~\cite{Brown:1990ev}. In the limit of infinite fermion masses the theory approaches the Yang-Mills theory (the quenched limit), that undergoes a genuine first-order transition. One hence expects to find a $\mathbb{Z}_2$ critical point between the region where the phase transition is a crossover and of the first order. The value of this critical point has been estimated on the lattice using the hopping parameter expansion~\cite{Fromm:2011qi,Saito:2011fs,Aarts:2016qrv} and via direct estimation using dynamical Wilson fermions and reweighting~\cite{Alexandrou:1998wv,Cuteri:2020yke}. There are also lattice-informed calculations of the thermodynamic equation of state~\cite{Ratti:2005jh,Philipsen:2012nu}.

Pioneering studies of the finite-temperature behaviour of ${\rm SU}(4)$ theories coupled to matter fields~\cite{LatticeStrongDynamics:2020jwi,Ayyar:2024dmt,Ayyar:2025eim,Ayyar:2026wht} and ${\rm SU}(3)$ theories with a large number of fermions, $N_f$~\cite{Schaich:2015psa}, provide indications of a  behaviour similar to that of QCD. Yet, for ${\rm Sp}(2N)$  theories coupled to fundamental matter, with $N>1$, no numerical study has been undertaken so far. Qualitatively, one expects the theory to undergo a first-order phase transition in the large-mass regime. As the mass is lowered, the transition is expected to weaken until a critical point is reached, below which it gives way to a smooth crossover.
Given the interest in these theories as models of dark matter, determining the critical parameters, and characterising the strength of the transition in the whole parameter space is of broad relevance to model builders and phenomenologists, and this paper presents the results of the first steps in this direction, for ${\rm Sp}(4)$ coupled to $N_{\rm f}=2$ Dirac fermions transforming in the fundamental representation.

\section{Lattice field theory}
\label{Sec:lattice}

In this section, we describe the lattice formulation of the continuum theory introduced in  Sect.~\ref{Sec:continuum}, presenting our choice of  discretised action,  the observables of interest, the calibration exercises we performed on the lattice ensembles, and the numerical analysis we implemented in order to study the  finite temperature regime of the lattice field theory.

\subsection{Lattice action and ensembles}
\label{Sec:LatticeActionAndEnsembleGeneration}

We study the lattice theory defined by a hypercubic lattice with isotropic spacing, $a$. The spacetime lattice volume is given by $\tilde{V} = a^4 (N_t\times N_s^3)$, with $N_t$ and $N_s$ the number of lattice points in the temporal and spatial dimensions, respectively.  Each lattice point is labelled by an integer four-vector, $ x \equiv (n_t, n_x, n_y, n_z)$. The link variable, $U_\mu( x)$, where $\mu$ labels the four spacetime components of the link,  is an element of the gauge group, and acts as the gauge connection between $x$ and nearest neighbour points. The lattice action, $S$,  is the sum of two separate terms, a gauge action, $S_g$, and a fermion action, $S_f$,
\begin{equation}
    S \equiv S_g + S_f\,,
    \label{eq:TotalLatticeAction}
\end{equation}
where both contributions are obtained by summing over the whole discretised space. As prescribed in Section~\ref{Sec:finiteT},  bosons have periodic boundary conditions in every direction, while the fermions  have periodic boundaries in the spatial dimensions and antiperiodic in the temporal dimension.

For the purposes of this  exploratory work, we adopt the standard Wilson Euclidean action~\cite{Wilson:1974sk}, defined in terms of the plaquette operator:
\begin{equation}
    P_{\mu\nu} (x) \equiv U_\mu(x) 
    U_\nu(x + \hat{\mu}) 
    U_\mu^\dagger(x + \hat{\nu})
    U_\nu^\dagger(x)\,,
    \label{eq:PlaquetteOperator}
\end{equation}
where $\hat{\mu}$ and $\hat{\nu}$ are unit vectors along the directions of the hypercubic lattice.
The Wilson gauge action for ${\rm Sp}(2N)$ is expressed in terms of the real part of the trace of the plaquette operator, summed over the spacetime:
\begin{equation}
    S_g = \beta \sum_{x} \sum_{\mu<\nu}
    \Big{(}
    1 - \frac{1}{2N} \mathrm{Re} \,\mathrm{Tr} \,P_{\mu\nu} (x)
    \Big{)}\,,
    \label{eq:LatticeGaugeAction}
\end{equation}
where $\beta \equiv \frac{4N}{g_0^2}$ is the lattice coupling,  and $g_0$ the bare coupling. This paper focuses on the ${\rm Sp}(4)$ group ($N=2$).

The $N_{\rm f}=2$ fermions, $Q^J$, transforming in the fundamental representation, appear in the Wilson-Dirac action:
\begin{equation}
    S_f = a^4
    \sum_{x}
    \sum_{J=1}^{N_f} \overline{Q^J} D_m Q^J\,,
    \label{eq:LatticeFermionAction}
\end{equation}
where spinor and gauge indices are understood. The action of the Wilson-Dirac operator, $D_m$, is defined as follows:
\begin{equation}
    \begin{aligned}
    D_m Q(x) \equiv& \left( \frac{4}{a} + m_0 \right)  Q(x) 
   - \frac{1}{2a} \sum_{\mu}\Big{[}
        (1-\gamma_\mu)U_\mu(x) Q(x + \hat{\mu})
       +
       (1+\gamma_\mu)U_\mu^\dagger(x-\hat{\mu}) Q (x-\hat{\mu})
    \Big{]} \,.
    \end{aligned} 
    \label{eq:WilsonDiracOperator}
\end{equation}

In order to  estimate numerically the observables, we replace the ensemble average in Eq.~(\ref{Eq:generalObservable}) with a finite sample on the lattice, generated using the (Hamiltonian) HMC algorithm~\cite{Duane:1987de}---see also Ref.~\cite{DelDebbio:2008zf}. We use a second-order Omelyan integrator~\cite{Takaishi:2005tz} to implement the molecular dynamics steps, without preconditioning the fermion matrix~\cite{DeGrand:1990dk}. For the implementation of the HMC we use the Grid software library~\cite{Boyle:2015tjk,Boyle:2016lbp,Yamaguchi:2022feu,Grid-repo}. Grid is a highly optimised, portable C++ library for simulating lattice gauge theories, adaptable to different architectures, including multi-GPU systems. Our implementation and testing of the ${\rm Sp}(4)$ theory in Grid are described in Ref.~\cite{Bennett:2023gbe}. In order to handle effectively the large number of ensembles required for this work, some of us developed a thin wrapper library for Grid, named hmcdj~\cite{hmcdj-repo}, which has the express purpose of managing large parameter scans~\cite{hmcdj}.

\subsection{Observables and the phase transition}

The first observable of interest  is the average plaquette, $P$, which for ${\rm Sp} (2N)$ gauge theories is defined as the spacetime and direction average:
\begin{equation}
    P \equiv 
    \frac{a^4}{6 \tilde{V}} \sum_x \sum_{\mu<\nu} 
   \Big{(}
     \frac{1}{2N} \mathrm{Re} \,\mathrm{Tr} \,P_{\mu\nu} (x)
    \Big{)}\,,
\label{eq:AveragePlaquette}
\end{equation}
with $P_{\mu\nu}$ the plaquette operator in Eq.~(\ref{eq:PlaquetteOperator}). The average plaquette is related to the gauge contribution to the Euclidean lattice gauge action, and the associated energy, $E$, by Eq.~(\ref{eq:LatticeGaugeAction}). We rewrite that relation as follows: 
\begin{equation}
    \label{eq:EuclideanLatticeGaugeAction}
    E \equiv \frac{1}{\beta} S_g = \frac{6 \tilde{V}}{a^4}\left(1\frac{}{}-P\right)\,,
\end{equation}
as this form is useful to discuss reweighting.
Notice that $E$ does not include the contribution of fermions.

The primary observable in our analysis is the lattice version, $\Phi$, of the Polyakov loop, defined in Eq.~(\ref{eq:ContinuumPolyakovLoop}). This is the order parameter of the deconfinement phase transition in the Yang-Mills theory. As one expects the contribution of fermions to decouple in the limit of parametrically heavy fermion masses (the regime in which the quenched approximation holds) we use the Polyakov loop as a proxy for an order parameter of the thermal transition also in the presence of finite fermion masses, despite the explicit symmetry breaking effects introduced by dynamical fermions~\cite{Cuteri:2020yke}. 
For ${\rm Sp}(2N)$ lattice gauge theories, the Polyakov loop  is defined as the trace over the product of all the temporal links, evaluated at a specific choice of spatial coordinates, $\vec{x}$:
\begin{equation}
    \Phi(\vec{x}) \equiv \frac{1}{2 N}
    \mathrm{Tr}
    \left[
    \prod_{t=0}^{N_t-1}
    U_0(t, \vec{x})
    \right]\,.
    \label{eq:PolyakovLoop}
\end{equation}
The average over spatial lattice sites of the Polyakov loop is then 
\begin{equation}
    \Phi \equiv \frac{1}{N_s^3} \sum_{\vec{x}} \Phi(\vec{x})\ .
    \label{eq:AveragePolyakovLoop}
\end{equation}

If the centre symmetry is exact, as in Yang-Mills theories,  the ensemble average of the Polyakov loop vanishes identically. Yet, the ensemble average of its absolute value, $\left|\Phi\right|$, defined by the relation
\begin{equation}
    \left|\Phi\right| = \frac{1}{N_s^3} \left|\sum_{\vec{x}} \Phi(\vec{x})\right|\,,
    \label{eq:AbsoluteAveragePolyakovLoop}
\end{equation}
  acts as an order parameter, vanishing in the low-temperature, confined phase, and being 
  non-zero in the high-temperature, deconfined phase, as discussed in Sect.~\ref{Sec:finiteT} for the continuum theory. We adopt this definition also for the theory coupled to matter fields, in which case the symmetry is not exact. It is convenient to determine the order of the phase transition by measuring a closely related observable,  the susceptiblity, $\chi_{\left|\Phi\right|}$, of the Polyakov loop, defined as
\begin{equation}
    \chi_{\left|\Phi\right|} \equiv
    N_s^3 \bigg(
    \Big\langle \left|\Phi\right|^2 \Big\rangle - \Big\langle \left|\Phi\right|\Big\rangle^2
    \bigg) \,.
    \label{eq:PolyakovSusceptibility}
\end{equation}

The conventional normalisation in Eq.~(\ref{eq:PolyakovSusceptibility}), proportional to the spatial volume, ensures that the Polyakov loop susceptibility, $\chi_{\left|\Phi\right|}$, is an extensive quantity, as typical of response functions. The peak of $\chi_{\left|\Phi\right|}$ is located at the (pseudo)critical value, $\beta_c$,  of the lattice gauge coupling. The study of the finite-volume scaling of the height of the peak is a standard diagnostic tool, used to inspect the order of a phase transition (see, e.g., Refs~\cite{Borsanyi:2012xf,Aarts:2014nba}, and references therein). The use of finite volume scaling to probe the order of a phase transition originates in pioneering work by Binder and Landau~\cite{Binder:1984llk}. In the case of a first-order phase transition, the order parameter is discontinuous across the transition, and its susceptibility has a $\delta$-function non-analyticity. In the discretised (regularised) version of the system provided by the lattice formulation, the discontinuity of the order parameter is smoothened out. Yet, in approaching the infinite spatial volume limit, $N_s\to\infty$, one expects a progressive steepening of the behaviour, with the value at the peak scaling linearly with the spatial volume:
\begin{equation}
    \chi^{\rm max}_{\left|\Phi\right|} \sim N_s^3\,.
\end{equation}
At the other extreme, if the height of the peak of the susceptibility tends to a constant with increasing spatial volume, this is interpreted as a smooth crossover:
\begin{equation}
    \label{eq:PolyakovSusceptibilityCrossoverScaling}
    \lim_{N_s\to\infty} \chi^{\rm max}_{\left|\Phi\right|}(N_s) \to \chi^{\rm max}_{\left|\Phi\right|} < \infty\,.
\end{equation}
The position of the peak is used to define  a (pseudo)critical coupling, $\beta_c$, even when the Polyakov loop is not an exact order parameter~\cite{Alexandrou:1998wv,Cuteri:2020yke}, in the presence of fermions.  We will test the scaling of $\chi^{\rm max}_{\left|\Phi\right|} $ numerically, later in the body of the paper, exploiting multiple histogram reweighting, which we discuss in details in Section~\ref{Sec:Reweighting}.

\subsection{Zero-temperature calibration and scale setting}
\label{Sec:Scale}

In this subsection, we present results of tests we perform on the lattice field theory at zero temperature. We aim at characterising in physical terms the regions of lattice parameter space we explore. We hence  measure the light meson spectroscopy and the Wilson flow scale. We also want to check that the statistical properties of the ensembles we generate with our numerical algorithms do not signal the presence of large systematic effects or pathologies invalidating our analysis, such as the breakdown of ergodicity. To this purpose, we monitor the topology of the ensembles, compute the autocorrelation and verify that all ensembles are thermalised.

All the ensembles described in this subsection have been generated using the Grid environment~\cite{Boyle:2015tjk,Boyle:2016lbp,Yamaguchi:2022feu,Grid-repo}, with the ${\rm Sp}(4)$ implementation described in Ref.~\cite{Bennett:2023gbe}, using machines with GPU architecture, and managed with  the hmcdj package~\cite{hmcdj,hmcdj-repo}. The Wilson flow computations have also been performed through Grid, while the correlators have been extracted from the configurations using Hadrons~\cite{antonin_portelli_2023_8023716, Hadrons-repo}.  The bulk phase structure of the lattice theory has been studied in~\cite{Bennett:2017kga}, and all measurements have been performed in the region of lattice parameter space connected to the continuum theory of interest.

\subsubsection{Meson masses}
\label{Sec:masses}

In order to set the scale of dimensionful quantities, and at the same time characterise the physical regime identified by the values of the lattice parameters  at which the thermal transition (or crossover) takes place, we find it useful to compute the zero-temperature spectrum of the lightest mesons. To this purpose, and anticipating our finite temperature results in Section~\ref{Sec:Results}, we choose values of the lattice coupling, $\beta$, and Wilson fermion mass, $a m_0$, in proximity to the (pseudo)transition, estimated from the Polyakov loop susceptibility peaks, $\chi^{\rm max}_{\left|\Phi\right|}(N_s)$.  

We generate new ensembles in which the lattice extends to large number of sites over all spacetime direction, with lattice volume $\tilde{V}=a^4(48\times 24^3)$, and extract the masses from the measurement of two-point correlation functions. As we find that  $N_s a m_{PS} > 7.5$, these volumes are large
enough that finite volume effects can be neglected, compared with the statistical uncertaintes, as suggested for this theory in Ref.~\cite{Bennett:2019jzz}.

 \begin{table}[t]
    \centering
    \caption{ The values of the lightest flavoured  pseudoscalar (${\rm PS}$) and vector (${\rm V}$) meson masses, in the ${\rm Sp}(4)$ theory coupled to $N_{\rm f}=2$ Dirac fermions transforming in the fundamental representation.  Measurements are performed in ensembles representative of the zero temperature limit, on lattices of dimension $48 \times 24^3$. The lattice action parameters, $\beta$ and $a m_0$, are chosen in proximity to the transition identified in the finite temperature numerical studies discussed later in the paper, for $N_t=4$ (top section of the table) and $N_t=6$ (bottom). We report the masses both in lattice units and in units of the Wilson flow scale, $w_0$.  The ratio $m_{\rm PS}/m_{\rm V}$ provides an indication of how far the masses are from the limit of massless fermions.}
    \vspace{5mm}
    \begin{tabular}{|c|c|c|c|c|c|c|c|}
\hline
$am_0$ & $\beta$ & $am_{\mathrm{PS}}$ & $am_{\mathrm{V}}$ & $w_0 / a$ & $ w_0 m_{\mathrm{PS}} $ & $ w_0 m_{\mathrm{V}} $ & $m_{\mathrm{PS}} / m_{\mathrm{V}}$ \\ 
\hline
\hline
$4$ & $7.32$ & $3.9883(5)$ & $3.9892(5)$ & $0.8134(3)$ & $3.244(1)$ & $3.245(1)$ & $0.9998(2)$ \\ 
$2.5$ & $7.31$ & $3.4540(9)$ & $3.4547(9)$ & $0.8041(2)$ & $2.7775(10)$ & $2.7780(10)$ & $0.9998(4)$ \\ 
$-0.6$ & $6.92$ & $1.434(3)$ & $1.475(4)$ & $0.58688(8)$ & $0.842(2)$ & $0.866(2)$ & $0.972(3)$ \\ 
$-0.87$ & $6.72$ & $1.168(3)$ & $1.248(6)$ & $0.54846(5)$ & $0.641(2)$ & $0.685(3)$ & $0.936(5)$ \\ 
\hline
\hline
$-0.3$ & $7.38$ & $1.334(4)$ & $1.350(4)$ & $1.1285(5)$ & $1.505(4)$ & $1.523(5)$ & $0.988(4)$ \\ 
$-0.5$ & $7.3$ & $1.062(5)$ & $1.091(6)$ & $1.1094(6)$ & $1.178(5)$ & $1.210(7)$ & $0.973(7)$ \\ 
$-0.65$ & $7.17$ & $0.893(6)$ & $0.942(9)$ & $1.0084(4)$ & $0.901(6)$ & $0.950(9)$ & $0.95(1)$ \\ 
$-0.77$ & $7.05$ & $0.750(6)$ & $0.83(1)$ & $0.9376(4)$ & $0.703(5)$ & $0.78(1)$ & $0.91(1)$ \\ 
$-0.87$ & $6.95$ & $0.562(9)$ & $0.67(2)$ & $0.9418(5)$ & $0.529(8)$ & $0.63(2)$ & $0.83(3)$ \\ 
$-0.92$ & $6.85$ & $0.642(8)$ & $0.77(2)$ & $0.7842(3)$ & $0.504(6)$ & $0.60(2)$ & $0.84(3)$ \\ 
\hline
\end{tabular}

    \label{table:scale_setting}
\end{table}

We perform calculations of flavoured pseudoscalar and vector masses corresponding to the operators in Table~\ref{tab:mesons}.  We define the zero-momentum two-point correlation function of the meson, ${\cal O}_M$, as follows
\begin{equation}
    C_{\mathcal O_M}(t) \equiv 
    \sum_{\vec{x}} \bra{0}
    \mathcal O_M(t,\,\vec{x})
    \mathcal O_M^\dagger(0,\,\vec0)
    \ket{0} \,,
    \label{eq:OperatorCorrelator}
\end{equation}
and we implement the source and sink operators as stochastic $Z_2$ wall sources~\cite{Boyle:2008rh}.

At large Euclidean time, the correlation functions are dominated by the lowest lying energy eigenstates with the same quantum numbers as the interpolating operators. The zero-momentum correlation functions take the form:
\begin{equation}
    \lim_{t\to\infty} C_{\mathcal O_M}(t) 
    \rightarrow
    \bra{0} \mathcal O _M \ket{M}
    \bra{0} \mathcal O _M \ket{M}^*
    \frac{1}{2m_M} \left[
    e^{-m_M t} + e^{-m_M(N_t a-t)} \right] \,,
    \label{eq:CorrelatorLargeTimeLimit}
\end{equation}
where $N_t a$ is the extent of the temporal direction. For the limited purposes of this work, as we perform only a small number of such measurements, we identify the plateaux and extract the masses with a two-parameter fit of the correlation function.
The fitting ranges are available in the data and workflow releases accompanying this publication~\cite{workflowrelease}. Our numerical results for the spectrum are reported in Table~\ref{table:scale_setting}. All the ensembles have large fermion masses, close to the quenched limit, as $m_{\rm PS}/m_{\rm V}\sim 1$. Some of the ensembles have large masses in lattice units $a m_{\rm PS},\,a m_{\rm V}\gsim 1$, and are hence affected by potentially large cutoff effects, due to the finite lattice spacing.

\begin{figure}[t]
    \centering
    \includegraphics{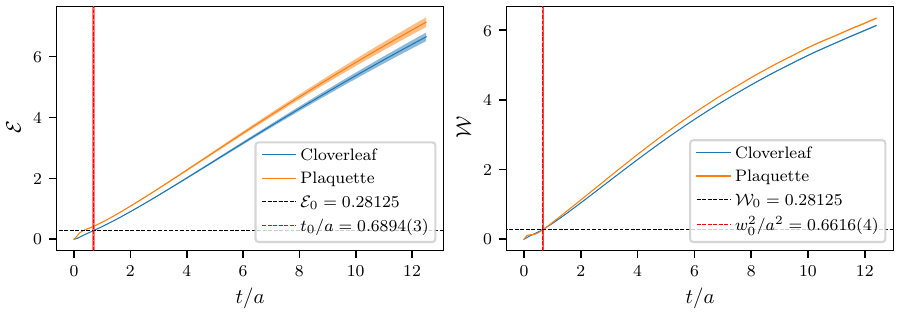}\\
    \includegraphics{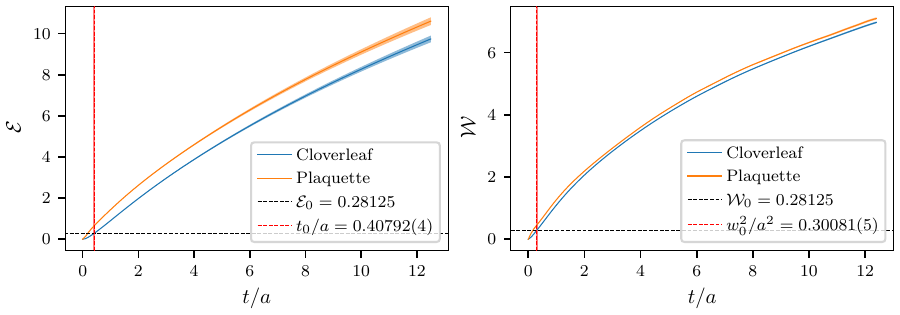}
    \caption{Representative examples of the Wilson flow, used for scale setting purposes, for one heavy (top panels, lattice parameters  $am_0=4$, $\beta=7.32$) and one light (bottom, $am_0=-0.87$, $\beta=6.72$) ensembles. In both cases, the lattice volume is $\tilde{V}=a^4 (48 \times 24^3)$. On the left panels, we illustrate the estimate of $t_0$, based on reference values of $\mathcal E_0$, and compare the result of using either  the plaquette in Eq.~(\ref{eq:PlaquetteOperator}), or the cloverleaf in Eq.~(\ref{eq:CloverLeafPlaquette}), as the discretisations of $G_{\mu\nu}$, to showcase the effect of the choice of discretisation in the Wilson flow action. The right panels show our extraction of $w_0$, with the conventional choice by the TELOS collaboration to adopt the reference value $\mathcal W_0 = 0.28125$~\cite{Bennett:2024tex, Bennett:2024cqv}.
    \label{Fig:ScaleSetting}}
\end{figure}

\subsubsection{Gradient flow and Wilson flow scale}
\label{Sec:GF}

\begin{figure}[ht]
    \centering
    \includegraphics{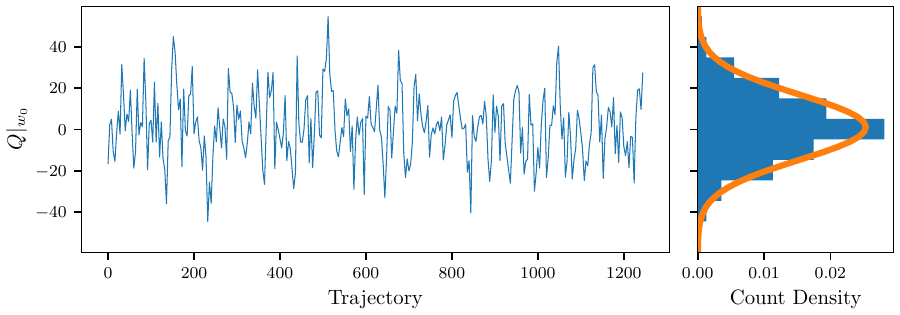}
    \includegraphics{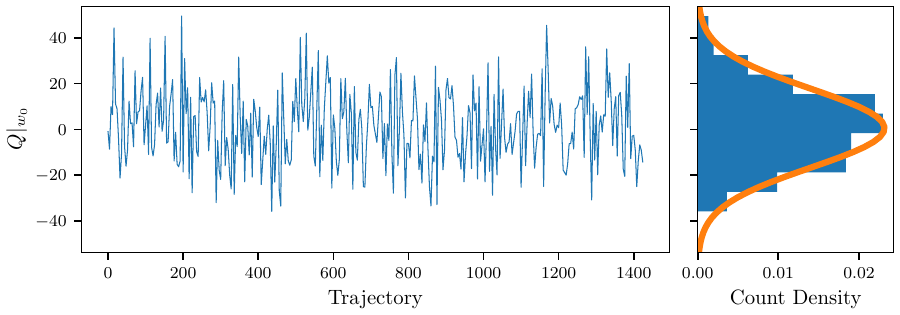}
    \caption{Representative examples of the Monte Carlo histories (left panels) and histogram distributions (right) of the topological charge calculated at a flow time equal to the scale $w_0^2$, $Q|_{w_0}$,  for one heavy (top panels, lattice parameters  $am_0=4$, $\beta=7.32$) and one light (bottom, $am_0=-0.87$, $\beta=6.72$) ensemble. In both cases, the lattice volume is $\tilde{V}=a^4 (48 \times 24^3)$. A Gaussian fit is superimposed to the histograms.   
    }
    \label{Fig:TopologicalCharge}
\end{figure}

We elect to use the gradient flow~\cite{Luscher:2010iy}, and its lattice implementation as the Wilson flow~\cite{Luscher:2013vga}, as our choice of scale setting procedure, allowing us to convert the dimensionless results of lattice calculations into physical units. The gradient flowed gauge field, $B_\mu(x,t)$, is introduced as the solution of the flow equation:
\begin{equation}
    \frac{\mathrm d B_\mu (t, x)}{\mathrm d t} =
    D_\nu G_{\nu\mu}(t, x)\,,
    \label{eq:WilsonFlow}
\end{equation}
where $t$ is in this subsection a fictitious flow time, $D_\nu$ is the covariant derivative,
 $G_{\mu\nu} = \left[ D_\mu, D_\nu \right]$ is the field-strength tensor as discretised using the Wilson gauge action.
 The flow equation is solved subject to the boundary condition dictated by $A_\mu(x)$, the gauge bosons:
\begin{equation}
    B_\mu(0, x) = A_\mu(x)\,.
    \label{eq:WilsonFlowInitialCondition}
\end{equation}

Following the conventions in Refs.~\cite{Bennett:2017kga, Bennett:2022ftz}, we define the dimensionless quantities, $\mathcal E(t)$ and $\mathcal W(t)$:
\begin{equation}
    \mathcal E(t)\ \equiv \ 
    \frac{t^2}{2}
    \left\langle       \frac{}{}
    \mathrm{Tr}
    \left[
     G_{\mu\nu}(t, x)
     G_{\mu\nu}(t, x)
     \right]        \frac{}{}
     \right\rangle\,,
    \label{eq:DimensionlessWilsonFlowEObservable}
\end{equation}
and
 \begin{equation}
     \mathcal W(t)\ \equiv \ 
     t \frac{\mathrm{d}}{\mathrm d t}
     \mathcal E(t)\,.
     \label{eq:WilsonFlowWObservable}
 \end{equation}
We then evolve these quantities along the flow, until reaching reference values. Common choices are the scale $t_0$,
\begin{equation}
    \mathcal E|_{t=t_0} \equiv \mathcal E_0\,,
\end{equation}
and the scale $w_0$~\cite{BMW:2012hcm},
 \begin{equation}
     \mathcal W|_{t=w_0^2} \equiv \mathcal W_0\,,
     \label{eq:DefW0}
 \end{equation}
 both of which rely on conventional choices of the reference values, $\mathcal E_0$ and $\mathcal W_0$. Throughout this work, we use the definition $\mathcal W_0 = 0.28125$, as chosen by the TELOS Collaboration for the Sp(4) theory~\cite{Bennett:2024tex, Bennett:2024cqv} on the basis of the expected quadratic Casimir scaling.

On the lattice, the discretised version of Eq.~(\ref{eq:WilsonFlow}) is solved by replacing $A_\mu(x)$ in Eq.~(\ref{eq:WilsonFlowInitialCondition}) with the link variable, $U_\mu(x)$. Thus $B_\mu(x,t)$ is replaced with $V_\mu(x,t)$, which obeys the initial condition:
\begin{equation}
    \label{eq:GradientFlowInitialCondition}
    V_\mu(x, 0) = U_\mu(x)\,.
\end{equation}
Different discretisations of $G_{\mu\nu}$ can be used in~\eqref{eq:DimensionlessWilsonFlowEObservable} to calculate $\mathcal{E}(t)$. The naive choice is to adopt the replacement with the plaquette operator $\mathcal P_{\mu\nu}$, defined in Eq.~(\ref{eq:PlaquetteOperator}). Alternatively, we can use the cloverleaf plaquette operator, $\mathcal C_{\mu\nu}$, which is defined as~\cite{Sheikholeslami:1985ij,Hasenbusch:2002ai}:
\begin{equation}
\begin{aligned}
    \label{eq:CloverLeafPlaquette}
    \mathcal C_{\mu\nu}(x) \equiv
    \frac{1}{2} 
    \bigg[\ 
    &U_\mu(x) U_\nu(x + \hat\mu) U_\mu^\dagger (x+\hat\nu) U_\nu^\dagger(x) \ + \\
    &U_\nu(x) U^\dagger_\mu(x+\hat\nu-\hat\mu) U_\nu^\dagger(x-\hat\mu) U_\mu(x-\hat\mu) \ + \\
    &U^\dagger_\mu(x-\hat\mu) U^\dagger_\nu(x-\hat\nu-\hat\mu) U_\mu(x-\hat\nu -\hat\mu) U_\nu(x-\hat\nu) \ + \\
    &U^\dagger_\nu(x-\hat\nu) U_\mu(x-\hat\nu) U_\nu(x-\hat\nu +\hat\mu) U_\mu^\dagger(x) \ 
    -\ \mathrm{h.c.}\bigg]\,.
\end{aligned}
\end{equation}
We show some representative choices of the Wilson flow, which include comparing the two choices of discretisation, in Fig.~\ref{Fig:ScaleSetting}. We report the measurements of the scale $w_0$ in
 Table~\ref{table:scale_setting}, for zero-temperature ensembles generated with choices of the lattice parameters close to the phase transitions identified for ensembles with $N_t=4$ and $N_t=6$.

\subsubsection{Topological charge}

In order to ensure ergodicity in our Monte Carlo, we monitor that our ensembles do not suffer from topological freezing, a phenomenon in which the sampling algorithm exhibits such large autocorrelation time for topological observables that the topology does not effectively change. The gradient flow acts as a smoothening procedure, removing short-distance flutuations, such that at a given flow time, $t$, the gauge field is smoothed over a finite radius, $\sqrt{8t}$. We combine this observation with the discretised lattice definition of the topological charge, which we write as follows:
\begin{equation}
    Q(t) = \frac{1}{32\pi^2}\sum_i
    \epsilon^{\mu\nu\rho\sigma}\ 
    \mathrm{Tr} \left[
    \mathcal C_{\mu\nu}(x,t) \mathcal C_{\rho\sigma}(x,t)
    \right]\,,
    \label{eq:TopChargeLattice}
\end{equation}
where $\mathcal C_{\mu\nu} (x, t)$ is the clover-leaf plaquette operator, defined in Eq.~(\ref{eq:CloverLeafPlaquette}). In Fig.~\ref{Fig:ScaleSetting} we show explicitly the dependence on the flow time, $t$,  after which the topological charge is computed.
We show representative examples of the Monte Carlo histories and histograms of the topological charge in Fig.~\ref{Fig:TopologicalCharge}. As the examples illustrate, we do not find significant evidence of topological freezing, and the distributions of topological charges is compatible with a Gaussian distribution.

\subsection{Density of states and reweighting}
\label{Sec:Reweighting}

We recall here the treatment of lattice observables using the density of states and reweighting. These are well established techniques, and the material presented in this subsection can be found in pedagogical introductions to the field, such as Ref.~\cite{Newman1998-tj}. Nevertheless, we find it useful to provide sufficient details to make the presentation self-contained and to fix the notation, paying special attention to conventional choices of signs and normalisations. We also highlight the limitations inherent to  histogram reweighting, which play an important role in Sect.~\ref{Sec:Results}.

In a quantum field theory that depends just on one control parameter, $\beta$, the path integral over all field configurations can be rewritten in terms of the density of states, denoted  as $\rho(E)$, which effectively counts the number of states with Euclidean action given by $\beta E$.  The partition function of the continuum theory,  ${\cal Z}(\beta)$, reads as follows:
\begin{equation}
    {\cal Z}(\beta) \equiv \int \mathrm{d}E \,\rho(E) e^{-\beta E}\,,
    ~\label{eq:SRw-partition-function-continuum}
\end{equation}
where $\beta$ is the lattice coupling.
For any operator, ${\cal O}(E)$,  the value of which depends only on $E$, the associated observable, computed with the path integral, can be recast as a one-dimensional integral:
\beqs
\left\langle {\cal O} \right\rangle_{\beta} &\equiv& \frac{1}{{\cal Z}(\beta)} \int \mathrm{d}E\, {\cal O}(E) \rho(E) e^{-\beta E}\,.
\label{eq:vevofO}
\eeqs

Computing the density of states, $\rho(E)$,  is tantamount to solving the theory, as all observables can then be computed with elementary integrals that make use of $\rho(E)$.\footnote{In the LLR algorithm one approximates the logarithm of $\rho(E)$ as a piecewise linear function, the coefficients of which are computed by an iterative stochastic procedure~\cite{Langfeld:2012ah, Langfeld:2013xbf,Langfeld:2015fua,Cossu:2021bgn}, which can be applied to Yang-Mills lattice gauge theories~\cite{Springer:2021liy,Mason:2022trc,Mason:2022aka,Springer:2023wok, Lucini:2023irm,Mason:2023ixv,Bennett:2024bhy,Mason:2024dve,Bennett:2025whm,Zierler:2026hyt}.}  Reweighting~\cite{Ferrenberg:1988yz,Ferrenberg:1989ui} is a numerical analysis technique that complements Monte-Carlo sampling methods, by taking advantage of the formulation of the path integral in terms of the density of states, but without attempting to compute $\rho(E)$ explicitly.  It is widely used in lattice field theory---see, e.g., Refs.~\cite{Csikor:1998eu,Alexandrou:1998wv,Fodor:2001au,Csikor:2004ik,Fodor:2004nz} and the more recent examples in Refs.~\cite{Saito:2011fs,Fromm:2011qi,Borsanyi:2016ksw,Jahn:2018dke,Brandt:2019ttv,Ejiri:2019csa,Bachtis:2020dmf,Cuteri:2020yke,Kiyohara:2021smr,Borsanyi:2024xrx}. Its purpose is to optimise the efficiency with which the physical information contained in ensembles, generated with sampling methods, is combined and analysed, hence reducing numerical costs. Doing so is particularly important in regions of parameter space in  proximity  to a phase transition. Given a calculation performed with a discrete set of values for the control parameters, reweighting allows us to extend the calculation of the observables for other nearby values of the input parameters.

Reweighting is introduced by noting that, given a real function, $f(x)$, and a known probability density function, $p(x)$, that is non-zero over the whole support of $f$, which we call $D$, definite integrals of $f(x)$ can be rewritten as:
\begin{equation}
    \int_D \mathrm{d}x\,f(x) =
    \int_D \mathrm{d}x\ p(x)\,\frac{f(x)}{p(x)} \equiv
    \left\langle \frac{f(x)}{p(x)} \right\rangle_{p}\,.
    \label{eq:Srw-estimator-definition}
\end{equation}
In the last expression, $p(x)$ has become part of the measure defining the ensemble average.  We describe in the following how to make use of this observation in the context of the canonical formulation of lattice field theory. First, we replace the integrals with summations extending over all configurations in the discretised lattice version of spacetime. Second, we approximate these summations with  sums involving samples, governed by the underlying probability distributions of the Monte Carlo calculations. To this purpose, we introduce the concept of an estimator. Following the pedagogical  presentation in Ref.~\cite{Newman1998-tj}, we start with the case of a single histogram~\cite{Ferrenberg:1988yz}, and generalise to multiple histograms~\cite{Ferrenberg:1989ui}. For the purposes of this work some of us developed an open source, light-weight Python package, using just-in-time compiled core functions, called \texttt{pyRw}~\cite{pyRw}, that implements single and multiple histogram reweighting. 

The canonical formulation of lattice field theory is closely related to the canonical ensemble in statistical mechanics. 
By discretising the spacetime, the partition function, ${Z}(\beta)$, and all observables are computed as ensemble averages, given by weighted summations over the space of configurations. As long as $\beta$ is the only control parameter in the system, one rewrites the ensemble average by replacing the sum over configurations with a summation over the microscopic states, with weight function, $e^{-\beta E}$, controlled by the Euclidean action, $\beta E$.   The partition function, $Z(\beta)$, is rewritten as
\begin{equation}
    Z(\beta) \equiv \sum_{E}N(E) e^{-\beta E}\,,
     \label{eq:SRw-partition-function-definition}
\end{equation}
where $N(E)$ enumerates the states with given value of $E$, and is the discretised version of the density of states, $\rho(E)$. The probability mass function,  $p_\beta(E)$,  is defined as
\begin{equation}
    p_\beta(E) \equiv \frac{1}{Z(\beta)}N(E) e^{-\beta E}\,.
    \label{eq:SRw-canonical-pdf-definition}
\end{equation}
Hence, the expectation value of an observable $O=O(E)$, which depends on the microscopic energy $E$, can be computed for any given value of $\beta$ as
\beqs
\left\langle O\right\rangle_{\beta} &=& \sum_E p_{\beta} (E) O(E) \,=\, \frac{1}{{Z}(\beta)}\sum_E O(E) N(E) e^{-\beta E}\,,
\label{Eq:latticeobservable}
\eeqs
The observables of the continuum theory, $\left\langle {\cal O}\right\rangle_{\beta} $, are recovered by taking the infinite volume and continuum limits of the lattice ones, $\left\langle O\right\rangle_{\beta}$.

In the expression of ${ Z}(\beta)$ in  Eq.~(\ref{eq:SRw-partition-function-definition}), and $p_\beta(E)$ in Eq.~(\ref{eq:SRw-canonical-pdf-definition}), $N(E)$ is independent of $\beta$. By inverting Eq.~(\ref{eq:SRw-canonical-pdf-definition}) to extract $N(E)$, and replacing the result in the definition of $p_{\beta^{\prime}}(E)$, for $\beta^{\prime}\neq \beta$, one arrives at the following expression:
\begin{equation}
    p_{\beta} (E) = 
    \frac{p_{\beta^{\prime}}(E)e^{(\beta^{\prime}-\beta)E}}
    {\sum_{E^{\prime}}p_{\beta^{\prime}}(E^{\prime})
    e^{(\beta^{\prime}-\beta)E^{\prime}}}\,,
    \label{eq:Srw-extrapolated-pdf}
\end{equation}
which relies only on the assumptions that  $N(E)\neq 0$. For values of $E$ for which both $p_{\beta} (E) $ and $p_{\beta^{\prime}} (E)$ are non-zero, this exact expression relates the probability distributions computed for two different values of the coupling.

\subsubsection{Monte Carlo estimators and single histogram reweighting}
\label{Sec:SingleHistogramReweighting}

A Monte Carlo estimator is a stochastic way of computing the summation in the right-hand-side of  Eq.~(\ref{Eq:latticeobservable})---and eventually the integral in Eq.~(\ref{eq:vevofO})---by estimating it with a finite sampling, governed by a known probability distribution, $p(E)$, that has support in the whole space of configurations for which $O(E)$ is non-trivial:
 \begin{equation}
\left\langle { O}\right\rangle_{\beta} \,=\,
    \left\langle \ 
    { O}(E)\,
    \frac{e^{-\beta E}}{p( E)} 
    \ \right\rangle_p\,.
    \label{eq:Srw-estimator-numerator}
\end{equation}
Given an ensemble of $N$ configurations, distributed according to  $p(E)$,
 Eq.~(\ref{eq:Srw-estimator-numerator}) is approximated by  the quantity
\begin{equation}
O_N \,\equiv\, \frac{\sum_{i=1}^N O_i 
    \frac{\exp(-\beta E_i)}{p_i}}
    {\sum_{j=1}^N \frac{\exp(-\beta E_j)}{p_j}}\,\simeq\,  \left\langle {\cal O}\right\rangle_{\beta} \,,
    \label{eq:Srw-importance-sampling-estimator}
\end{equation}
where $O_i$, $p_i$, and $\beta E_i$ are, respectively,  the value of operator, probability, and  Euclidean action, evaluated on the $i$-th configuration in the ensemble. The relation in Eq.~(\ref{eq:Srw-importance-sampling-estimator})  is exact in the limit $N\rightarrow +\infty$.
Hence,  it provides a good approximation provided the number of configurations, $N$, is large enough.
The estimator, $O_N$, reduces to a simple arithmetic average if one sets $p_i = \frac{1}{Z(\beta)}\exp(-\beta E_i)$, reproducing the distribution of typical sampling algorithms.

Having sampled the distribution for a value of the coupling, $\beta^{\prime}\neq\beta$, different from that of interest,  one can choose
\begin{equation}
    p_i = \frac{1}{Z(\beta^{\prime})}\exp(-\beta^{\prime} E_i)\,,
    \label{eq:Srw-interpolating-pdf}
\end{equation}
and by replacing this expression for $p_i$ in Eq.~(\ref{eq:Srw-importance-sampling-estimator}),  after simplifying the $Z(\beta^{\prime})$ factors, one arrives at
\begin{equation}
    O_N = \frac{\sum_{i=1}^N O_i 
    e^{(\beta^{\prime}-\beta) E_i}}
    {\sum_{j=1}^N e^{(\beta^{\prime}-\beta) E_j}}\, ,
    \label{eq:SingleHistogramReweighting}
\end{equation}
with the observable $O$ being esimated at the target value $\beta$ and the measurements $O_i$ being sourced from the ensemble generated at $\beta^\prime$.
This estimator is used in numerical calculation, and captures the exact property of
the probability distributions in Eq.~(\ref{eq:Srw-extrapolated-pdf}), relating the probability distributions with different coupling, $\beta\neq \beta^{\prime}$.  

The necessary condition for applicability of Eq.~(\ref{eq:Srw-estimator-definition}) is that $p(x)$ should not vanish inside the support of $f(x)$, otherwise the formal manipulation leading to the use of the estimator breaks down.  
The version of this statement adapted to the case of an estimator built from a finite Monte Carlo sample, as in Eq.~(\ref{eq:SingleHistogramReweighting}), can be formulated by requiring that the histogram in the known energy distribution, obtained with coupling $\beta^{\prime}$, must have a sizeable overlap with the one expected  with coupling $\beta$. 
For the purposes of this study, we impose the following  heuristic requirement:
\begin{equation}
    |E(\beta^{\prime}) - E(\beta)| \le \sigma_E\,,
    \label{ineq:Srw-valid-range}
\end{equation}
where $\sigma_E$ is the standard deviation of the distribution of values of $E$, sampled with coupling  $\beta^{\prime}$. For large-scale numerical calculations, involving ensembles much larger than those produced for the purposes of the present study,  this requirement is conservative, and could be relaxed, given that Eq.~(\ref{eq:Srw-extrapolated-pdf}) is expected to hold in the limit $N\rightarrow \infty$.

If several control parameters, $\vec{\kappa}$, are present, and the Euclidean action of the system can be expressed as a sum over a number of associated observables, $\vec{v}$, with the replacement $-\beta E\rightarrow -S_E = \vec{\kappa} \cdot \vec{v}$, then Eq.~(\ref{eq:SingleHistogramReweighting}) generalises to
\begin{equation}
    \label{eq:MultiDimReweighting}
    O_N  =  \frac{\sum_i^N O_i e^{-(\vec{\kappa}^{\prime}-\vec{\kappa}) \cdot \vec{v}_i}}
{\sum_j^N e^{-(\vec{\kappa}^{\prime}-\vec{\kappa}) \cdot \vec{v}_j}}\,,
\end{equation}
where $\vec{\kappa}^{\prime}$ is the vector of reference values of the parameters, for which a Monte Carlo sample exists, and $\vec{v}_i$ is the $i$-th measurement of the observables characterising the system, while $O_i$ is the value of the observable of interest. Notice the conventional (and harmless) change of signs in the definition of the parameters.

\subsubsection{Multiple histogram reweighting}

Reweighting can be generalised  to combine together information coming from more than one sample. If the system is controlled by only one control parameter, which we identify with the lattice coupling, $\beta$, and we have $M$ available finite ensembles, with lattice coupling $\beta_j$, each consisting of $N_j$ configurations, each having Euclidean action $\beta_j E_{j,s_j}$, where  $j=0,\,\cdots,\,M-1$, and $s_j=0,\,\cdots\,N_j-1$, then the partition function, with coupling $\beta$, obeys the relation:
\begin{equation}
 Z(\beta) = \sum_{i=0}^{M-1}\sum_{s_i=0}^{N_i-1}
    \left(
    \sum_{j=0}^{M-1} \frac{N_j}{ Z_j} e^{(\beta-\beta_j)E_{i,s_i}}\right)^{-1}\,.
    \label{eq:mrw_Z}
\end{equation}
The coefficients, $Z_j$, have to be determined iteratively, to meet self-consistency requirements, $Z_j=Z(\beta_j)$, for each $j=0,\,\cdots,\,M-1$.
Once the  values of $Z_j$ are known,  the observable related to a generic operator, $O$, is given by
\begin{equation}
    \langle O(\beta)\rangle = 
    \frac{1}{Z(\beta)}
    \sum_{i=0}^{M-1}\sum_{s_i=0}^{N_i-1}
     \left(
        \frac{O_{i,s_i}}
{   
   \sum\limits_{j=0}^{M-1} \frac{N_j}{ Z_j} e^{(\beta-\beta_j)E_{i,s_i}}}\right)\,,
    \label{eq:MultiHistogramReweighing}
\end{equation}
where $O_{i,s_i}$ is the value of the observable in the $s_i$-th element of the $i$-th ensemble. A detailed derivation of these results is outside the scope of this work and we refer the reader to the relevant section in Ref.~\cite{Newman1998-tj}. Multiple histogram reweighting is affected by a limitation similar to the inequality~(\ref{ineq:Srw-valid-range}), originating in the requirement on the support of $p(x)$ appearing in Eq.~(\ref{eq:Srw-estimator-definition}). This may cause numerical challenges when approaching the thermodynamic limit, as the Euclidean action is proportional to the lattice volume, and hence at larger lattice volumes the measurements in $\beta$ need to become increasingly denser.

\section{Numerical Analysis and Results }
\label{Sec:Results}

The ${\rm Sp}(4)$ gauge theory coupled to fermion matter fields has not so far been studied at finite temperature and its finite temperature phase structure in the space of  bare lattice parameters is unknown. 

On the lattice, we identify the temperature, $T$, with the inverse of the size of the temporal direction,
\begin{equation}
    \label{eq:LatticeTemperature}
    T \equiv  \frac{1}{aN_t}\,,
\end{equation}
having chosen $N_t \ll N_s$, as  prescribed in Sect.~\ref{Sec:finiteT}. 

 The first step of our investigation consists of fixing $N_t \ll N_s$, and selecting a small set of values of interest for the bare mass, $a m_0$, of the $N_{\rm f}=2$ degenerate Wilson-Dirac fermions transforming in the fundamental representation. Holding these fixed, we scan over a range of lattice couplings, $\beta$, which is related to the inverse of $a$, and hence to the temperature, $T$, via Eq.~(\ref{eq:LatticeTemperature}). We chose to repeat the process for the  two (small) choices of temporal lattice extent,  $N_t=4$ and $N_t=6$.  In both cases, we have also sampled the Yang-Mills case using the HMC, in order to compare our results with the limiting case of $am_0\to+\infty$.\footnote{The HMC is inefficient in sampling the Yang-Mills theory, when compared, for example, to adopting a combination of heatbath and over-relaxation, or the LLR algorithms, due to the emergence of large autocorrelation times. This exercise is performed as a cross-check of our results, and we have verified that it agrees with results found in the literature using other algorithms, see Refs.~\cite{mason_HiRep_LLR_v1.1.0,Bruno:2024dha}.} 
 
We compute the ensemble average of the absolute value of the Polyakov loop and its susceptibility for the aforementioned choices of the lattice parameters. This process provides a first estimation of the (pseudo)critical values of the coupling, $\beta_c$, for each choice of $N_t$, $N_s$, and $am_0$.  A refinement of this estimate is obtained by applying multiple histogram reweighting to the Polyakov loop susceptibility, and performing a finite-volume scaling analysis. As we shall see, the analysis becomes more expensive and challenging as the lattice volume increases, which is an expected consequence of Eq.~\eqref{eq:EuclideanLatticeGaugeAction}. Hence, for the purposes of this exploratory work,  we perform the reweighting analysis only for ensembles with $N_t=4$

 All the ensembles included in this study are generated using the HMC implementation of Grid~\cite{Boyle:2015tjk,Boyle:2016lbp,Yamaguchi:2022feu,Grid-repo}, supplemented by the TELOS adaptations  for the $Sp(4)$ group~\cite{Bennett:2023gbe}, and hmcdj~\cite{hmcdj}, on GPU architectures. Detailed characterisations of the large number of ensembles produced in this study can be found in the associated data release~\cite{data_release}.

\begin{figure}[t]
    \centering
    
    \includegraphics{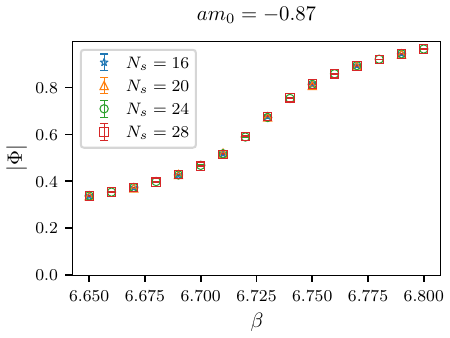}
    \includegraphics{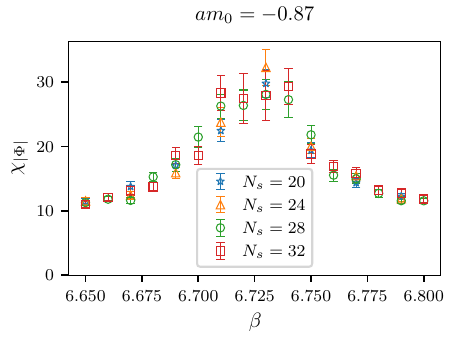}
    
    \includegraphics{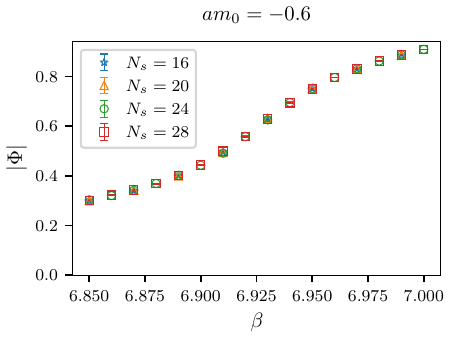}
    \includegraphics{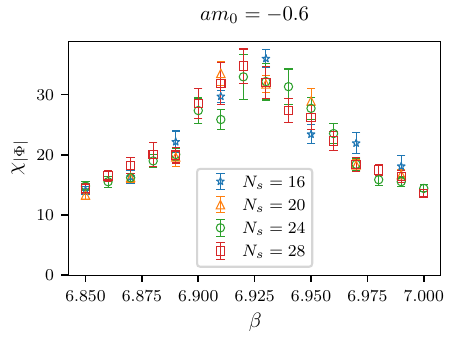}
    
    \includegraphics{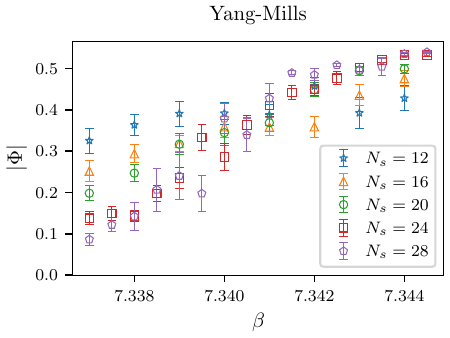}
    \includegraphics{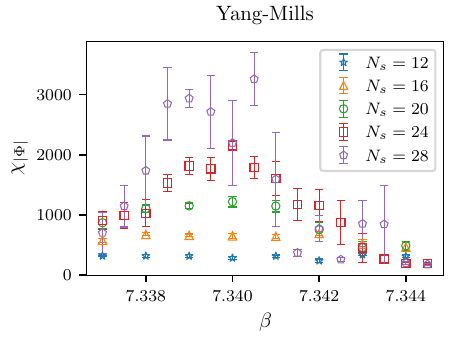}

    \caption{Ensemble average of the absolute value of the Polyakov loop, $|\Phi|$  (left panels), defined in Eq.~\eqref{eq:PolyakovLoop}, and the associated susceptibility, $\chi_{|\Phi|}$ (right), as a function of the lattice coupling, $\beta$, for two representative choices of bare mass, $am_0=-0.87$ (top panels) and $am_0=-0.6$ (middle), of the $N_{\rm f}=2$ Dirac fermions. The ensembles have fixed temporal extent, $N_t=4$, and we consider four choices of spatial extent, among $N_s=16,\, 20,\,24,\,28,\,32$.
   The two bottom panels show the same measurements, on lattices of temporal extent $N_t=4$, for the ${\rm Sp}(4)$  Yang-Mills theory, with five choices of spatial extent, $N_s=12,\,16,\,20,\, 24,\,28$.
    }
    \label{Fig:Nt4SparsePolyakovLoop}
\end{figure}

\begin{figure}[t]
    \centering
        \includegraphics{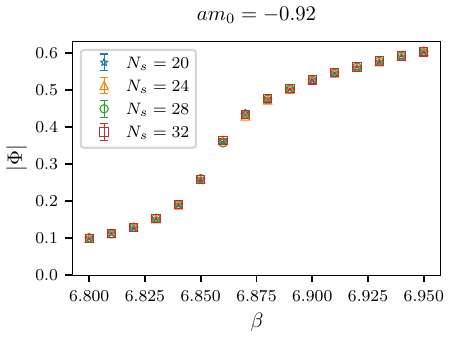}
        \includegraphics{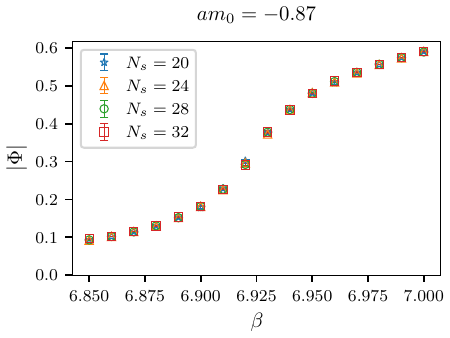}
        
        \includegraphics{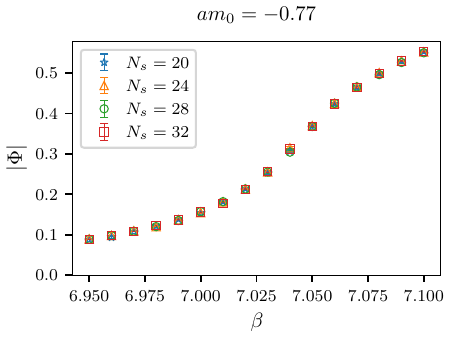}
        \includegraphics{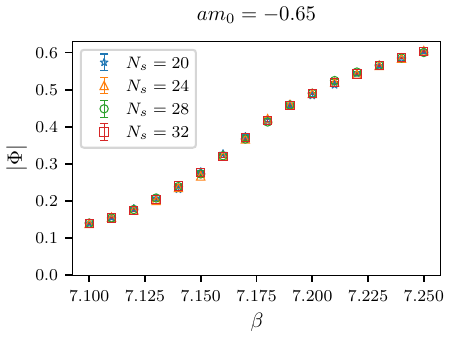}

        \includegraphics{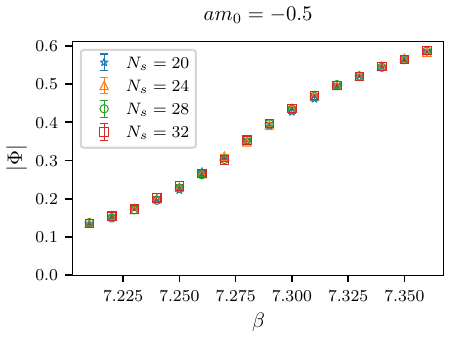}
        \includegraphics{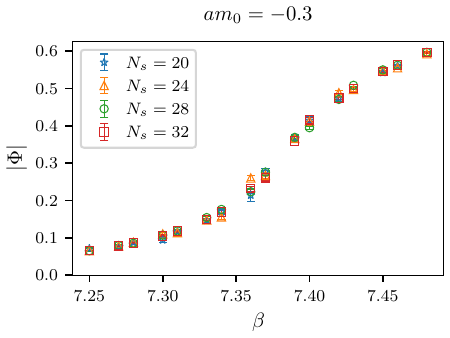}
    
    \caption{Ensemble average of the absolute value of the Polyakov loop, $|\Phi|$, defined in Eq.~\eqref{eq:PolyakovLoop}, as a function of the lattice coupling, $\beta$, in the ${\rm Sp}(4)$ theory coupled the $N_{\rm f}=2$ Dirac fermions with six representative choices of bare Wilson fermion mass, $a m_0=-0.97,\,-0.87,\,-0.77,\,-0.65,\,-0.5,\,-0.3$. The lattice  temporal extent is $N_t=6$, and different colours and markers represent several choices of spatial extent, $N_s=20,\,24,\,28,\,32$.}
    \label{Fig:Nt6SparsePolyakovLoop}
\end{figure}

\begin{figure}[t]
    \centering
    
        \includegraphics{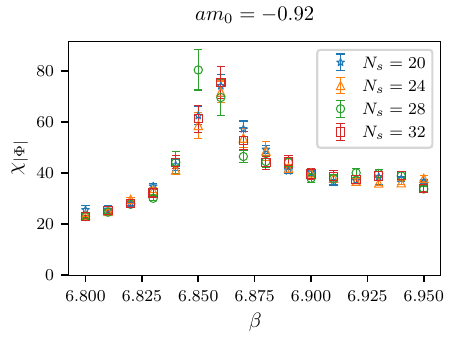}
        \includegraphics{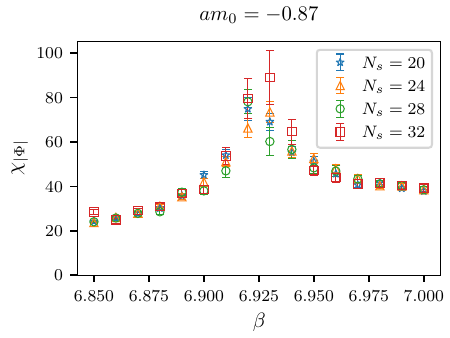}
        
        \includegraphics{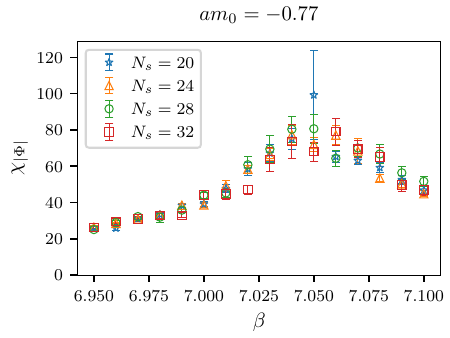}
        \includegraphics{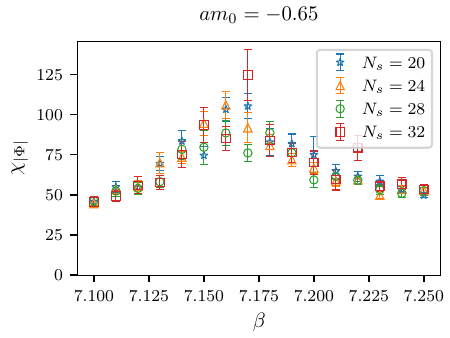}
        
        \includegraphics{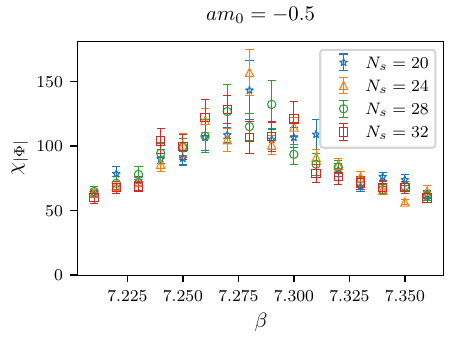}
        \includegraphics{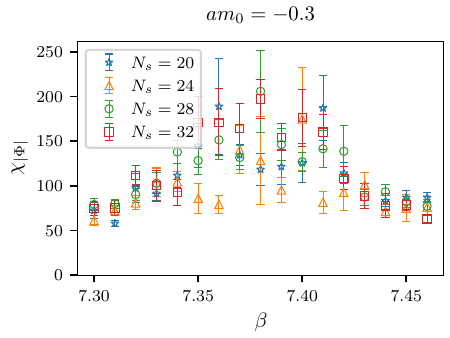}
        
    \caption{Susceptibility of the  absolute value of the Polyakov loop, $\chi_{\left|\Phi\right|}$, defined in Eq.~\eqref{eq:PolyakovSusceptibility}, as a function of the lattice coupling, $\beta$, in the ${\rm Sp}(4)$ theory coupled the $N_{\rm f}=2$ Dirac fermions, for six representative choices of bare Wilson fermion mass, $a m_0=-0.97,\,-0.87,\,-0.77,\,-0.65,\,-0.5,\,-0.3$. The lattice  temporal extent is $N_t=6$, and different colours and markers represent several choices of spatial extent, $N_s=20,\,24,\,28,\,32$.}
    \label{Fig:Nt6SparsePolyakovSusceptibility}
\end{figure}

\subsection{Polyakov loop, sparse scanning of lattice parameter space}

For the purpose of  identifying approximate values of the (pseudo)critical coupling, $\beta_c$, for a selection of values  of bare masses, $a m_0$, we measure the absolute value of the Polyakov loop, and its susceptibility, for two choices of temporal lattice extents, $N_t=4$ and $N_t=6$, and four different spatial extents, $N_s$.

\subsubsection{Small temporal direction, $N_t=4$}

The smallest temporal extent  we consider is $N_t=4$. We restrict attention to two values of the fermion mass, $am_0=-0.87$ and $am_0=-0.6$.  The ensemble average of the absolute value of the Polyakov loop is shown in Fig.~\ref{Fig:Nt4SparsePolyakovLoop}, together with its susceptibility, as a function of $\beta$. For both masses, we find clear evidence of a monotonically increasing Polyakov loop, and of  a peak in the susceptibility that does not scale in height and position with the spatial volume, $N_s^3$, of the lattice. The fact that all susceptibility curves  collapse on one another is the expected manifestation of a crossover.

As a consistency check, we perform the same exercise for the corresponding Yang-Mills theory, and show the results in the bottom panels of Fig.~\ref{Fig:Nt4SparsePolyakovLoop}. While the numerical results are noisy, due to the aforementioned inefficiency inherent in the use of the HMC algorithm, the results clearly show a different behaviour, in respect to the case of dynamical fermions.  The Polyakov loop average is still a monotonically increasing function of $\beta$, upon varying the spatial extent, $N_s$, the position of the peak shows a mild displacement, and its height a marked scaling. These are indications of a first-order phase transition separating the large-$\beta$ and small-$\beta$ regime of the lattice theory. We will return to these statements, and make them  more precise, by using multiple histogram reweighting, later in this section.

\begin{figure}[t]
    \centering
    \includegraphics{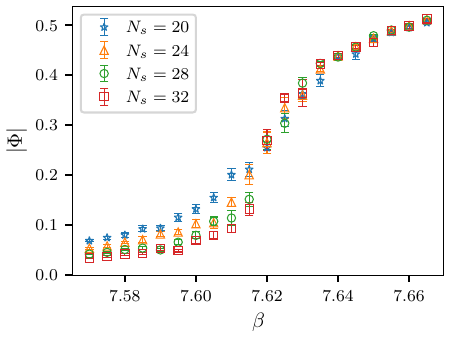}
    \includegraphics{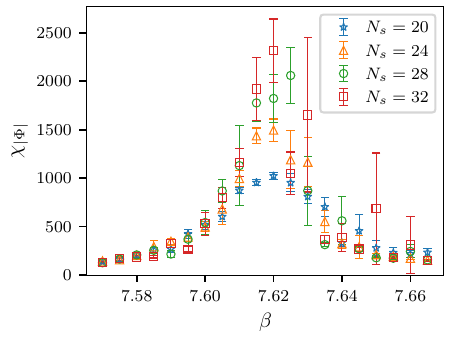}
    \caption{Ensemble average of the absolute value of the Polyakov loop, $|\Phi|$ (left panel), defined in Eq.~\eqref{eq:PolyakovLoop}, and its associated susceptibility, $\chi_{|\Phi|}$ (right), defined in Eq.~\eqref{eq:PolyakovSusceptibility}, in the ${\rm Sp}(4)$ Yang-Mills theory, plotted as a function of the lattice coupling, $\beta$, and measured  on lattices of temporal extent $N_t=6$, and spatial extents $N_s=20,\,24,\,28,\,32$.}
    \label{Fig:Nt6YangMills}
\end{figure}

\subsubsection{Towards the continuum limit: $N_t=6$}

The continuum limit, and the field theory, are recovered by repeating the exercise we performed for $N_t=4$ with increasingly larger values of $N_t$, paying attention to always take first the (thermodynamic) limit of infinite spatial volume, before extrapolating to large $N_t$. This is a gargantuan numerical endeavour, and before diving into it, we  check  that the expected behaviour appears when moving to $N_t=6$, and test our numerical analysis methodologies. Much of what we are going to say here has been tested at length in the context of QCD. Given the novelty of this exploratory study, these measurements are necessary, as a calibration for future large scale projects.

Hence, we repeat the exercises in the case  $N_t=6$. Guided by the findings  in the case of the Yang-Mills theory~\cite{Bruno:2024dha}, and as indicated by   Eq.~\eqref{eq:LatticeTemperature}, the (pseudo)critical values of the coupling, $\beta_c$, are expected to increase in respect  to $N_t=4$. Yet, qualitatively we should find a similar picture: a clear signal of a first-order phase transition for the Yang-Mills theory, and a crossover for small values of the fermion mass, $am_0$.

We perform a scan of the lattice  parameter space for six values of the bare fermion mass, in the range $-0.9\leq am_0 \leq -0.3$. The ensemble average of the absolute value of the Polyakov loop is shown in Fig.~\ref{Fig:Nt6SparsePolyakovLoop}, and the associated susceptibility in Fig.~\ref{Fig:Nt6SparsePolyakovSusceptibility}. We also analyse the Yang-Mills theory, using the HMC, in the vicinity of the (known) critical value, $\beta_c$, and the results are shown in Fig.~\ref{Fig:Nt6YangMills}.  We repeat our measurements for several choices of the spatial extent, $N_s$.

The results of the analysis are qualitatively the same as for the $N_t=4$ case. In the presence of (light) fermion masses, the location and height of the peak in the Polyakov loop susceptibility is independent of the spatial volume, indicating the presence of a crossover. By contrast, in the case of the Yang-Mills theory, we find that the Polyakov loop and its susceptibility exhibit scaling with the spatial volume, in agreement with the expectation that the Yang-Mills theory undergoes a first-order phase transition~\cite{Holland:2003kg, Mason:2023ixv}. We notice that the location of the peak agrees with the value of $\beta_c=7.624(5)$ found in the literature~\cite{Bruno:2024dha}, and obtained  with the heatbath and overelaxation algorithm.

\begin{figure}[t]
    \centering
    \includegraphics{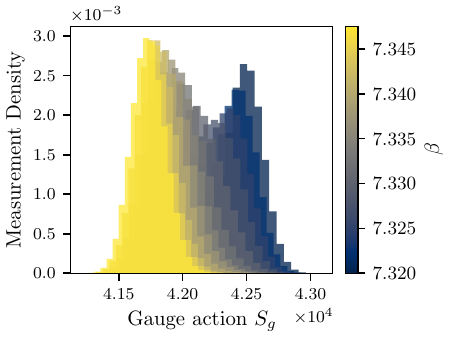}
    \includegraphics{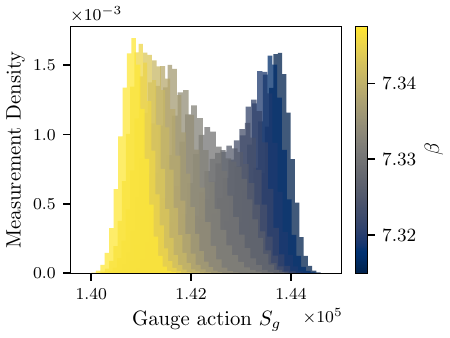}
    \caption{Representative examples of histograms of the gauge action, $S_g$, defined in Eq.~\eqref{eq:EuclideanLatticeGaugeAction}, computed in ensembles with temporal extent $N_t=4$, and spatial extent $N_s =16$ (left panel) and $N_s=24$ (right). In both cases, the fermion  mass is $am_0=4$, and we restrict attention to discrete choices of coupling,  $\beta$, in the critical region near $\beta_c$. The dense overlap of the histograms allows for the use of multiple histogram reweighting in the range of $\beta$s generated.
}
    \label{Fig:Nt4SparsePolyakovSusceptibility}
\end{figure}

\begin{figure}[t]
    \centering
    \includegraphics{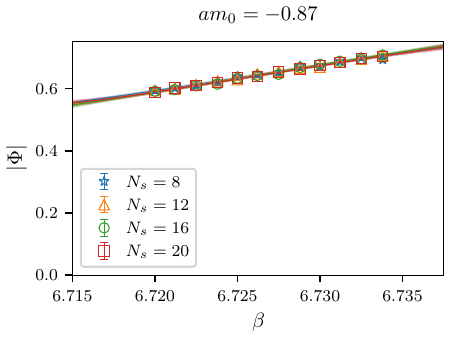}
    \includegraphics{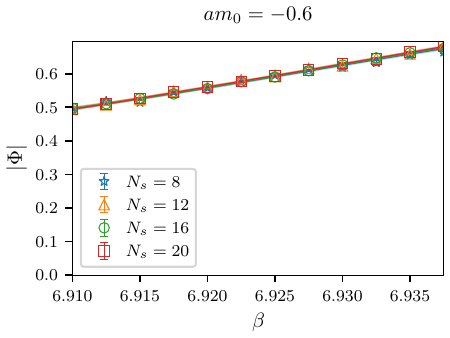}
    
    \includegraphics{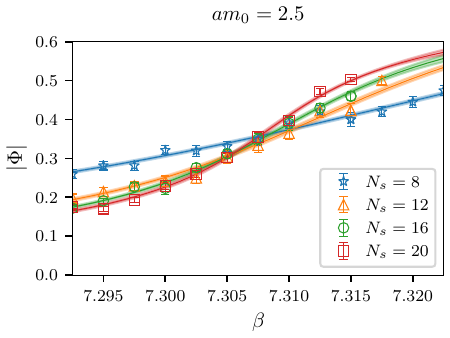}
    \includegraphics{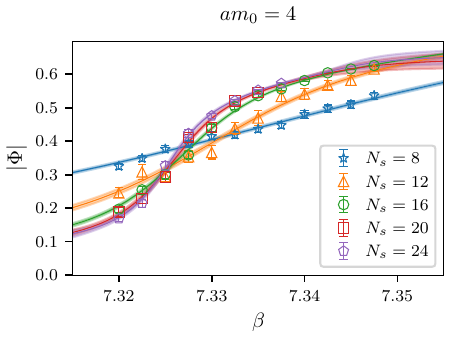}
    
    \includegraphics{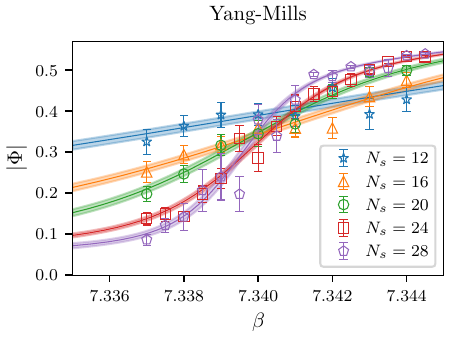}
    \caption{Ensemble average of the absolute value of the Polyakov loop, defined in Eq.~\eqref{eq:PolyakovLoop}, for lattices with temporal extent $N_t=4$, as a function of the coupling, $\beta$, analysed using reweighting (continuous curves). Ensembles with the two lightest masses, $am_0=-0.87$ and $am_0=-0.6$ (top panels), exhibit collapse of the curves obtained by changing the spatial extent, $N_s$. This is the expectation of smooth crossovers.    Those with the two heaviest masses, $am_0=2.5$ and $am_0=4$ (middle panels), exhibit scaling behaviour with the spatial volume,  suggestive of a first-order phase transition.  The behaviour of the Yang-Mills theory (bottom panel) is qualitatively similar to the heavy mass cases, though the signal is more clearly marked.}
    \label{Fig:Nt4ReweightingPolyakovLoop}
\end{figure}

\begin{figure}[t]
    \centering
    \includegraphics{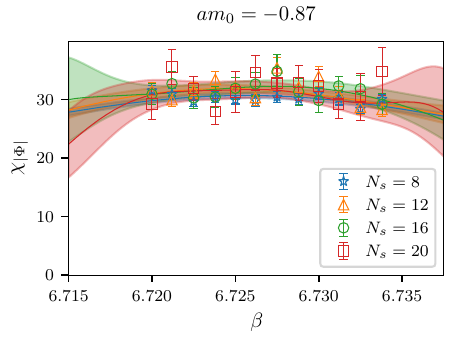}
    \includegraphics{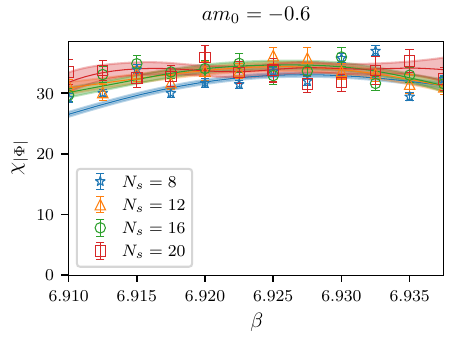}
    
    \includegraphics{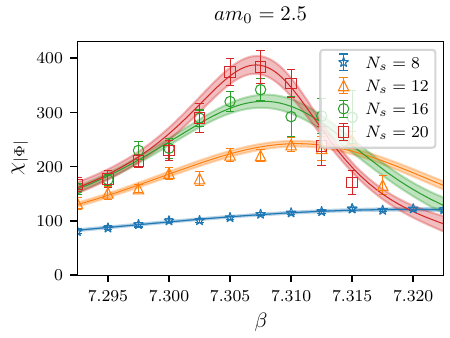}
    \includegraphics{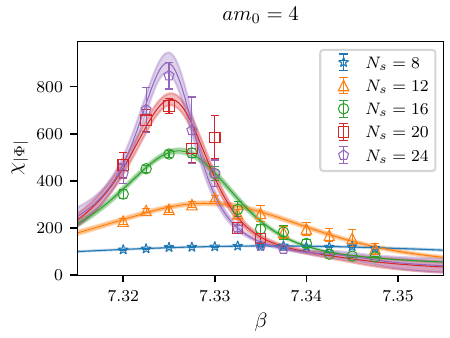}
    
    \includegraphics{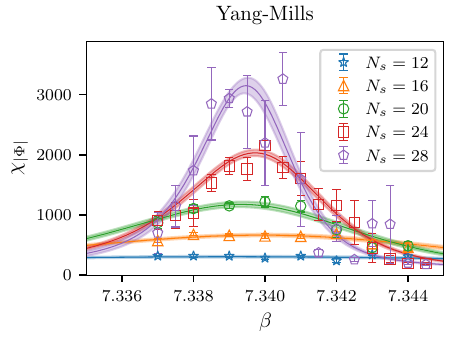}
  \caption{Susceptibility of the absolute value of the Polyakov loop, defined in Eq.~(\ref{eq:PolyakovSusceptibility}), for lattices with temporal extent $N_t=4$, as a function of the coupling, $\beta$, analysed using multiple histogram reweighting (continuous curves). Ensembles with the two lightest masses, $am_0=-0.87$ and $am_0=-0.6$ (top panels), exhibit collapse of the curves obtained by changing the spatial extent, $N_s$. This is the expectation of smooth crossovers.    Those with the two heaviest masses, $am_0=2.5$ and $am_0=4$ (middle panels), exhibit scaling behaviour with the spatial volume,  suggestive of a first-order phase transition.  The behaviour of the Yang-Mills theory (bottom panel) is qualitatively similar to the heavy mass cases.}
    \label{Fig:Nt4ReweightingPolyakovSusceptibility}
\end{figure}

\begin{figure}[t]
    \centering
        \includegraphics{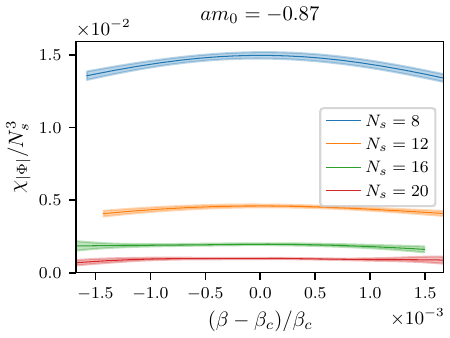}
        \includegraphics{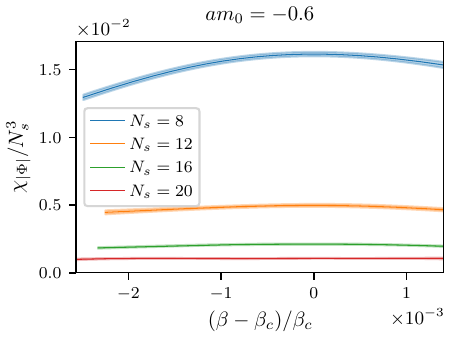}
        
        \includegraphics{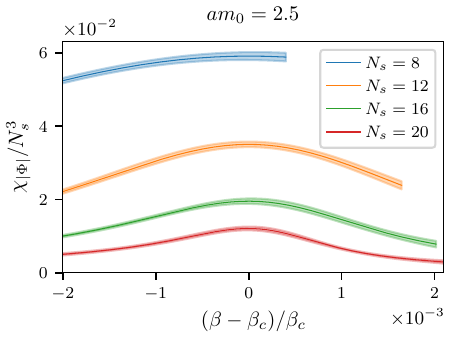}
        \includegraphics{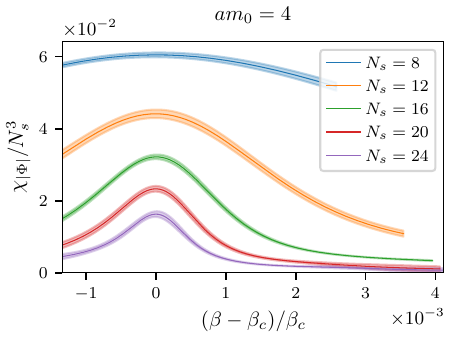}
        
        \includegraphics{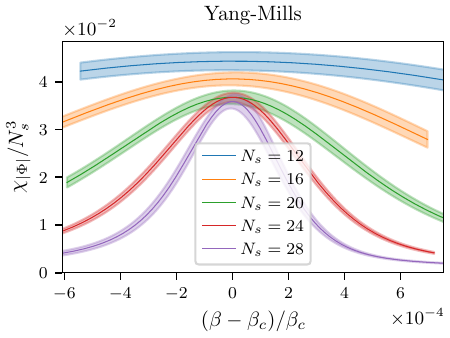}
    \caption{Results of the multiple histogram reweighting procedure, applied to the susceptibility of the absolute value of the Polyakov loop, $\chi_{|\Phi|}$, defined in Eq.~(\ref{eq:PolyakovSusceptibility}), and scaled by $N_s^3$, for the lattice ensembles generated with $N_t=4$ and for the same values of $N_s$, $am_0$, and $\beta$ as in Figs.~\ref{Fig:Nt4ReweightingPolyakovLoop} and~\ref{Fig:Nt4ReweightingPolyakovSusceptibility}. Note that the horizontal axis has been rescaled and centered around the estimated (pseudo)critical value $\beta_c$.}
    \label{Fig:Nt4ReweightingPolyakovSusceptibilityScaledByVolume}
\end{figure}

\subsection{Towards high precision: reweighting for $N_t=4$ }

Besides studying ensembles with larger $N_t$, we can improve our analysis by turning the discrete (sparse) scans presented so far into continuous curves that allow for precise determinations of the critical hyper-surfaces, and a better quality of the characterisation of the scaling with the volume. To this purpose, we apply multiple histogram reweighting (see Section~\ref{Sec:Reweighting}) to an extended set of ensembles with $N_t=4$. Given the evidence outlined earlier, we include in this exercise two additional choices of the fermion bare mass, $am_0=2.5$ and $a m_0=4$. These intentionally large numerical values allow us to gain access to the approach to the quenched limit, while still using dynamical Wilson fermions. 

We generate ensembles for a finite number of choices of lattice couplings, $\beta$, that lie in proximity to the (pseudo)critical value, $\beta_c$, for fixed fermion mass, $am_0$, and measure the Polyakov loop susceptibility. We then reconstruct the observable of interest for all values of $\beta$ close to the (pseudo)critical point using multiple histogram reweighting.  In order to estimate the uncertainties, we apply a bootstrapping procedure, based on binning the range of numerical measurements in intervals with width dictated by the integrated autocorrelation time, $\tint$:
\begin{equation}
    n = \frac{N}{\tint}\,,
    \label{eq:ReweigthingBootstrapBinning}
\end{equation}
with $N$ the number of  measurements of a given ensemble. We provide the definition of the integrated autocorrelation time in Eq.~(\ref{eq:IntergratedAutocorrelationTime}) and discuss its estimation in Appendix~\ref{App:Autocorrelation}. Each of the bootstrap elements is generated by random sampling (with replacement) of $n$ elements out of the $N$ original measurements and performing the reweighting on it.

We visually check that the histograms of the Wilson action, $S_g$, show clear evidence of overlap between ensembles generated with adjacent choices of coupling, $\beta$. As discussed in Section~\ref{Sec:Reweighting}, multiple histogram reweighting can be used effectively when the action histograms overlap. Representative examples  are displayed in Fig.~\ref{Fig:Nt4SparsePolyakovSusceptibility}. Larger couplings (larger temperatures) lead to larger values of the energy, as indicated by the color-coding in the histograms.  The goodness of the overlaps supports the reliability of the  multiple histogram reweighting process, within our accuracy and precision goals. We notice that an increase in lattice volume leads to sparser histograms, the individual distributions becoming narrower, so that a finer mesh in the choice of values for $\beta$ is needed to approach the thermodynamic limit.

Our main results are shown in Fig.~\ref{Fig:Nt4ReweightingPolyakovLoop}, for the ensemble average of the absolute value of the Polyakov loop, and in Fig.~\ref{Fig:Nt4ReweightingPolyakovSusceptibility}, for its susceptibility,  for four choices of fermion mass in the range $-0.87\leq am_0\leq 4$,  as well as for the Yang-Mills theory. In each case, we compare several choices of lattice spatial extent, $N_s$, while keeping fixed the temporal extent, $N_t=4$.

For the two lightest masses, which we have discussed in an earlier subsection, reweighting confirms that the Polyakov loop curves collapse on one another when measured with different spatial volumes, while the susceptiblity shows no sign of scaling with the increasing spatial volumes. We interpret this behaviour as confirmation that a crossover is taking place for these values of the lattice parameters.

In the case of the two heaviest bare masses available, $am_0=4$ and $am_0=2.5$,  the Polyakov loop curves computed for different spatial volumes and with the application of multiple histogram reweighting display the distinctive character expected for first-order phase transitions. In particular, the height of the peak of its susceptibility, $\chi_{|\Phi|}$ scales with the spatial volume, $V$. This is the first evidence in the literature of a  first-order thermal  phase transition in the ${\rm Sp}(4)$ theory coupled to massive fermion matter fields.
The behaviour of the heavy ensembles resembles qualitatively the one we find for the Yang-Mills theory, which is known to exhibit a first-order phase transition~\cite{Holland:2003kg,Mason:2023ixv,Bruno:2024dha}.

\begin{figure}[t]
    \centering
    \includegraphics{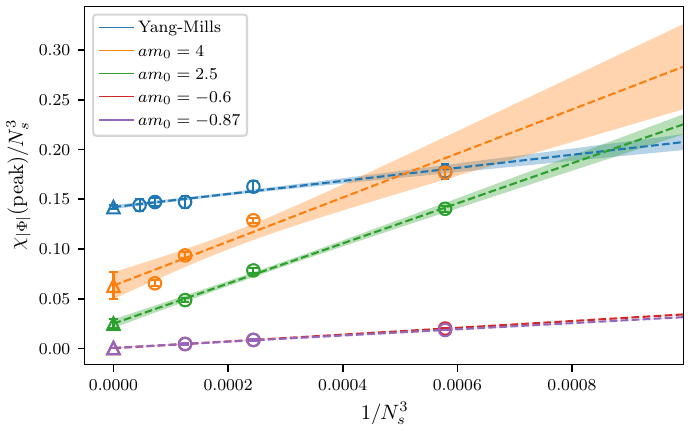}  
    \caption{Scaling of the ratios of the height of the peak of the Polyakov loop susceptibility over the spatial lattice extent, $\chi_{|\Phi|}^{\rm peak}/N_s^3$, in the case of lattices with temporal extent $N_t=4$, as a function of $1/N_s^3$,  for the same five sets of ensembles depicted in Figs.~\ref{Fig:Nt4ReweightingPolyakovLoop}, \ref{Fig:Nt4ReweightingPolyakovSusceptibility}, and~\ref{Fig:Nt4ReweightingPolyakovSusceptibilityScaledByVolume}.The data are represented by the circular points. The continuous curves and shaded bands correspond to the central values and error of a linear fit, and the values at $1/N_s^3\rightarrow 0$, marked by the triangular points, are the results of the infinite volume extrapolation. \label{Fig:Nt4Scaling}}
\end{figure}

In approaching the infinite volume (thermodynamic) limit, and in the presence of a first-order phase transition, one expects the height of the peak of the Polyakov loop susceptibility to scale proportionally to the volume, $V$. To verify this behaviour, we present in Fig.~\ref{Fig:Nt4ReweightingPolyakovSusceptibilityScaledByVolume} the dimensionless ratio $\chi_{|\Phi|}/N_s^3$, computed using the results of multiple histogram  reweighting, for the same lattice ensembles as in Figs.~\ref{Fig:Nt4ReweightingPolyakovLoop} and~\ref{Fig:Nt4ReweightingPolyakovSusceptibility}. We also shifted the value of coupling, $\beta$, to the critical value, $\beta_c$, hence removing the shift of the critical coupling due to finite volume effects.

Comparing the cases of light and heavy fermions  with the Yang-Mills case shows a marked difference, consistent with our interpretation that in the light cases we are studying a crossover, which turns into a first-order thermal transition for large masses, as in the Yang-Mills case.
The scaled plots in Fig.~\ref{Fig:Nt4ReweightingPolyakovSusceptibilityScaledByVolume} also show evidence of deviations from the expected scaling, especially for the smaller lattice volumes available. Nevertheless, while for the lightest masses there is no evidence of scaling, and hence the ratio $\chi_{|\Phi|}/N_s^3$ shows no particularly interesting features, in the case of the heaviest ensembles, as in the Yang-Mills case, for the largest available choices of $N_s$, the ratio, $\chi_{|\Phi|}/N_s^3$, tends to converge towards a common value near the peak. While these observations highlight the fact that spurious finite-volume effects are still present in many of our ensembles, yet the trend in the data strongly  supports the presence of a first-order transition for heavy masses.

In order to provide further quantitative support for this conclusion, we present linear fits of the peaks, for the four choices of mass, and the Yang-Mills theory alike, in Fig.~\ref{Fig:Nt4Scaling}.
In particular, we study the scaling of the peak value of the ratio $\chi_{|\Phi|}/N_s^3$ inspecting the infinite volume limit $\lim N_s^3\to\infty$. 
In qualitative agreement with our observations for Fig.~\ref{Fig:Nt4ReweightingPolyakovSusceptibility}, the ratio tends to a non-zero constant in the Yang-Mills and heavy fermion bare mass cases; the infinite volume limit of the ratio is compatible with zero in the case of light fermion mass.

\section{Outlook}
\label{Sec:outlook}

We have reported our numerical results from the first lattice study of the finite temperature behaviour of the ${\rm Sp}(4)$ theory coupled to matter fields. The matter content consists of $N_{\rm f}=2$ species of Dirac fermions transforming in the fundamental representation of the gauge group. The pseudo-real nature of this representation leads to symmetry enhancement effects. The corresponding, strongly coupled continuum theory is a prominent candidate for the short-distance origin of new physics models in the CHM and SIMP contexts.  

We have discretised the theory using the Wilson action coupled to $N_{\rm f}=2$ Wilson-Dirac fermions. Gauge configurations have been generated with the HMC algorithm, for lattices with isotropic spatial extent, $N_s a$, much larger than the size of the compact temporal direction $N_s \gg N_t $. We have scanned the lattice parameter space, by varying the fermion mass, $a m_0$, and the lattice coupling, $\beta$. We have computed the ensemble average of the absolute value of the Polyakov loop, $|\Phi|$, and its susceptibility, $\chi_{|\Phi|}$, to identifying the (pseudo)critical coupling, $\beta_c$, for $N_t=4$ and $N_t=6$. 

A finite-volume study of the observables, performed by repeating the analysis with increasing values of $N_s$, and measuring the scaling of the peak of the susceptibility, $\chi_{|\Phi|}$, with the lattice volume, provides a means of distinguishing between the presence of first-order phase transitions and smooth crossover. To this purpose, we have implemented the multiple histogram reweighting procedure, and successfully tested it against theoretical expectations, on ensembles with $N_t=4$, both for the well understood Yang-Mills theory (as a benchmark), as for the new cases with massive matter fields. We have also performed non-trivial calibration and scale setting exercises, by monitoring the Wilson flow and topology, and computing the masses of the lightest mesons in ensembles obtained in proximity to the critical values of the parameters with isotropic space-time, representative of the zero-temperature behaviour of the system. We have verified that our approach reproduces known results for the Yang-Mills theory, obtained with the alternative heatbath and LLR algorithms. More details of these final tests will be appearing in a future publication, currently in preparation~\cite{Bennett:2026algorithms}.

The extensive numerical exploration undertaken for this work (which produced in excess of five-hundred ensembles) has uncovered a rich phase structure in this lattice theory. We have presented strong evidence that, as for the Yang-Mills theory, in the presence of large fermion masses there is a first-order, finite-temperature  phase transition. The transition becomes weak and ultimately changes into a smooth cross-over for smaller values of the fermion mass. The picture that emerges, albeit preliminary and limited to ensembles with finite volume and finite lattice spacing, is compatible with expectations indirectly inferred from analogies with similar theories, such as QCD. In particular, a line of first-order phase transitions appears to be present at large values of the fermion mass, but terminates at a critical value, $am_0^{c}$, which is still not accurately known~\cite{Cuteri:2020yke}. In the process, we have also developed,  tested and benchmarked new software, implementing for ${\rm Sp}(4)$ the numerical algorithms needed for future large-scale studies of this and closely related theories with other matter content, or based on other gauge groups in the ${\rm Sp}(2N)$ family. As the ${\rm Sp}(4)$ theory provides the minimal realisation of the SIMP scenario for dark matter, the evidence we collected of a first-order transition in the presence of heavy matter fields represents the first successful milestone towards demonstrating the viability of gravitational wave experiments, such as the Einstein Telescope, as experimental probes of the origin of dark matter.

This work sets the stage for a plethora of future research projects. We provide an incomplete list of suggestions, ordered by increasing level of challenge. First, in order to capture the physics of the corresponding continuum theory, we would repeat similar calculations for larger values of $N_t$ (and, correspondingly, larger spatial volumes),  to perform the continuum-limit extrapolation. Second, we could study additional observables, for example the (screening) mass of the mesons at finite temperature, and verify whether the thermal transition is associated also with the restoration of the global symmetries acting on the fermions and bound states of the theory. Other observables, such the topological simplicity (introduced in Ref.~\cite{Crean:2026caz} and references therein) might help to characterise the transition. Third, a more systematic study of the parameter space would allow to identify the position of the critical point (even for $N_t=4$), and characterise it as a conformal field theory, in terms of its operators and their scaling dimensions. Last but not least, the pseudo-real nature of the fermion representations would allow the exploration of the phase structure in the presence of a finite chemical potential, as this special lattice theory is not affected by the sign problem. 

All these ambitious goals require the scaling up of resources for the numerical studies, and are hence deferred to the future. Their pursuit will also be aided by the development of bespoke tools and algorithms. In particular, it might be fruitful to adopt a different formulation of the lattice action, for example by using the M\"{o}bius implementation~\cite{Brower:2004xi,Brower:2005qw,Vranas:1997da,Vranas:2000tz,Berruto:2001ty,Brower:2012vk} of domain-wall fermions~\cite{Kaplan:1992bt,Shamir:1993zy,Furman:1994ky}, which has been recently tested for the ${\rm Sp}(4)$, $N_f=2$  theory at zero temperature~\cite{Bennett:2026cev}. This formulation has better global symmetry properties and built-in ${\cal O}(a^2)$ improvement~\cite{Bietenholz:1998ut}, compared to the ${\cal O}(a)$ of Wilson fermions~\cite{Sheikholeslami:1985ij}, yielding sizeable advantages in the approach to the continuum limit---see, for instance,  Ref.~\cite{Zhang:2026zpa} and references therein for recent progress on the lattice study of the phase diagram of QCD with domain wall fermions. The use of anisotropic lattices would also reduce the numerical cost of exploring the finite-temperature theory closer to the massless and continuum limits, as demonstrated in other contexts (see e.g., Refs.\cite{Klassen:1998ua,QCD-TARO:2000hup,Bolder:2000un,CP-PACS:2003ncs,Morrin:2006tf,Edwards:2008ja,Cheng:2010fe,Aarts:2015mma,Lee:2017uvl}). 

A very challenging goal would be the exploration in closer detail of the region of parameter space near criticality,  aiming to the study of phenomena related to phase coexistence and metastability. This will require the adoption of a different numerical strategy, for example by adapting the LLR algorithm beyond the Yang-Mills case~\cite{Bennett:2024bhy,Bennett:2025whm}, to work also in the presence of massive fermions. This challenging task would be linked to the contiguous programmes of study of out-of-equilibrium physics and bubble dynamics.  Such investigations would be of interest to a broad community, extending beyond lattice and high-energy phenomenologists, to encompass the astrophysics and cosmology community, in relation to the study of the primordial gravitational waves  discussed in the introduction. Prominently, a research programme of this type would yield reliable and testable estimates for the power spectrum of the relic density of gravitational waves. The  more formal community interested in developing applications of string theory and supergravity, in the form of gauge-gravity dualities, to the study of phase transition---see, e.g., Refs.~\cite{Bigazzi:2020phm,Ares:2020lbt,Bea:2021zsu, Bigazzi:2021ucw,Henriksson:2021zei,Ares:2021ntv,Ares:2021nap,Morgante:2022zvc,Bea:2021zol, Bea:2022mfb,Bea:2024xgv,Bea:2024bxu,Bea:2024bls,Hoyos:2026eoq,Bigazzi:2026ppu,Mateos:2026moc} and references therein---has also shown interest in lattice numerical results in these theories, as they can be compared to predictions that apply in the large-$N_c$ limit.

\begin{acknowledgments}

The work of A.~V.-P., B.~L., C.~A., E.~B., and M.~P. is supported by the STFC Consolidated Grant No.~ST/X000648/1. C.~A., E.~B., and M.~P. are supported by STFC [UKRI4243]. The work of E.~B. is supported by the STFC Research Software Engineering Fellowship EP/V052489/1 and has been supported in part by UKRI via the Computational Science Centre for Research Communities (CoSeC), under Grant No.~UKRI497. B.~L., L.~D.~D., and M.~P. received funding from the European Research Council (ERC) under the European Union's Horizon 2020 research and innovation program under Grant Agreement No.~813942. The work of L.~D.~D. and R.~H. is supported by the STFC Consolidated Grant No.~ST/X000494/1. The work of R.~H. is supported by the DiRAC 3 Operations Grant ST/X000079/1. D.~K.~H. is supported by Basic Science Research Program through NRF funded by the Ministry of Education (NRF-2017R1D1A1B06033701). J.-W.~L. is supported by IBS under the project code, IBS-R018-D1. C.-J.~D.~L. acknowledges support from NSTC Taiwan through grant number 112-2112-M-A49-021-MY3. C.-J.~D.~L. is also supported by the Taiwanese MoST Grant No.~109-2112-M-009-006-MY3. C.-J.~D.~L. is also supported by Grants No.~112-2639-M-002-006-ASP and No.~113-2119-M-007-013-. B. L. is supported in part by the STFC Consolidated Grants No. ST/X00063X/1 and No. UKRI4238. The work of D.~V. is supported by the STFC Consolidated Grant No.~ST/X000680/1 and UKRI/ST/C004237/1. The work of A.~V.-P. is supported by a studentship funded by the Faculty of Science and Engineering, Swansea University, and by the School of Physics and Astronomy, The University of Edinburgh.

This work used the DiRAC Data Intensive service (DIaL3) at the University of Leicester and Extreme Scaling service (Tursa) at the University of Edinburgh, managed by the University of Leicester Research Computing Service and EPCC respectively on behalf of the STFC DiRAC HPC Facility (www.dirac.ac.uk). The DiRAC service at Leicester and the DiRAC service at Edinburgh were funded by BEIS, UKRI and STFC capital funding and STFC operations grants. DiRAC is part of the UKRI Digital Research Infrastructure.

Numerical computations were performed using Supercomputing Wales SUNBIRD at Swansea University and AccelerateAI at Swansea University.

Supercomputing Wales and AccelerateAI are supported by the European Regional Development Fund via Welsh Government.

{\bf Research Software Availability statement}---Raw data were generated using Grid~\cite{Boyle:2015tjk,Boyle:2016lbp,Yamaguchi:2022feu,Grid-repo} commit \texttt{e809c66f}, found in the TELOS Collaboration's fork~\cite{TELOSGrid}, hmcdj~\cite{hmcdj, hmcdj-repo} commit \texttt{a06264f} and Hadrons~\cite{antonin_portelli_2023_8023716,Hadrons-repo} commit \texttt{f3fb3f9e}. The workflow used to analyse these data is available at Ref.~\cite{workflowrelease}.

{\bf Research Data Availability Statement}---The raw data generated in support of this work, and processed data derived from it, are available in machine-readable format at Ref.~\cite{datarelease}. See also Ref.~\cite{Bennett:2025neg}, for a description of our approach to reproducibility and open science. 

{\bf Open Access Statement}---For the purpose of open access, the authors have applied a Creative Commons  Attribution (CC BY) license to any Author Accepted Manuscript version arising.

\end{acknowledgments}

\appendix

\begin{figure}[t]
    \centering
    \includegraphics{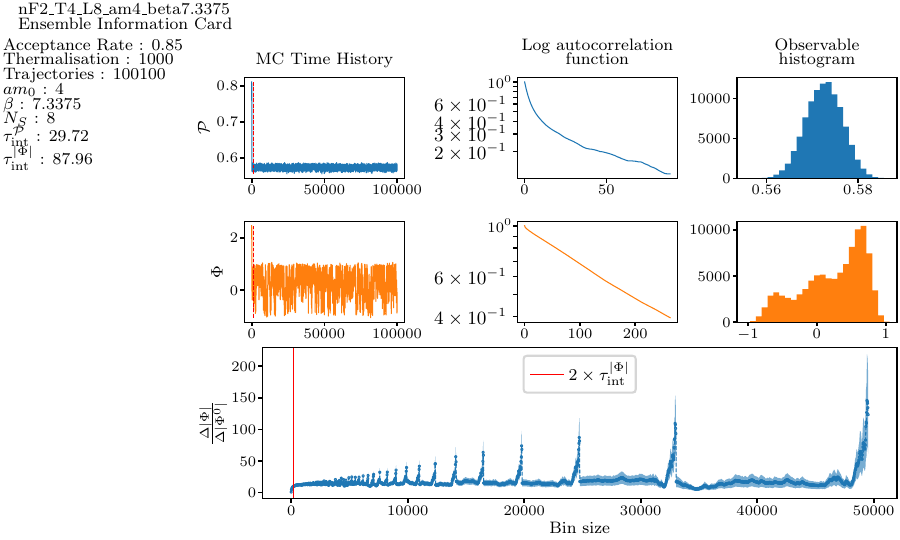}
    \includegraphics{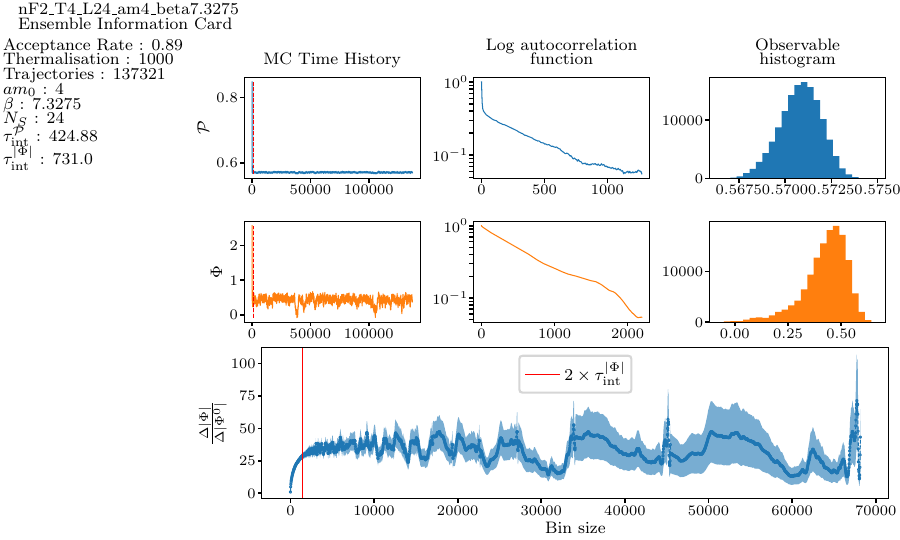}
    \caption{Ensemble information ``cards'' for two ensembles with the same $\beta = 7.3375$,  $N_t=4$, and $am_0=4$, and two different choices of spatial volume  $V=(8a)^3$ (top panel) and $V=(24a)^3$ (bottom). During production of HMC chains, we must monitor thermalisation and the level of autocorrelation in each ensemble. We monitor observables, collecting them in the ensemble cards presented. We also monitor the autocorrelation of the absolute value of the Polyakov loop by with the binning method and compare with the integrated autocorrelation time. We expect the variance ratio to plateau near the calculated value of $\tint$.}
    \label{Fig:EnsembleCards}
\end{figure}

\begin{figure}[t]
    \centering
    \includegraphics{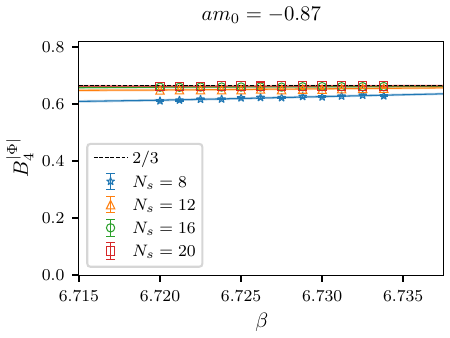}
    \includegraphics{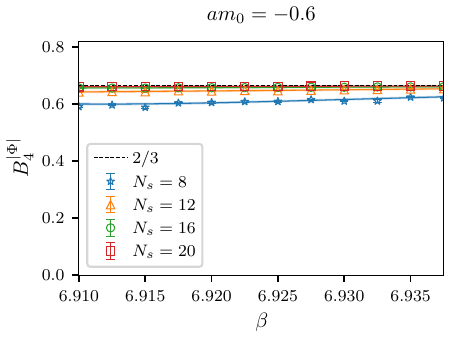}
    
    \includegraphics{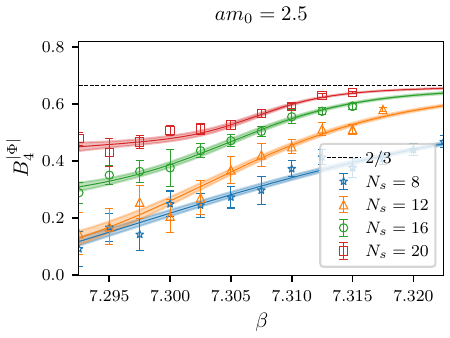}
    \includegraphics{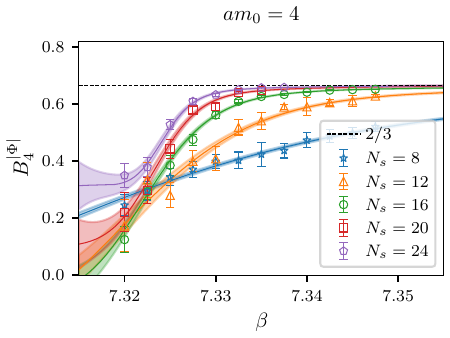}
    
    \includegraphics{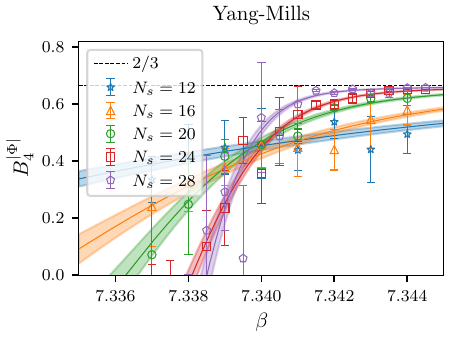}
    \caption{Binder cumulant, defined in Eq.~\eqref{eq:BinderCumulant},  for the Polyakov loop, $X=\Phi$, for discrete values of the coupling, $\beta$, and for fixed  lattice  temporal extent, $N_t=4$. The panels show the results of applying the multiple histogram reweighting to the ${\rm Sp}(4)$ theory coupled to two Wilson-Dirac fermions with mass $am_0$ (top and middle panels), and the Yang-Mills theory (bottom). Different colours and symbols denote different choices of spatial extent, $N_s$.}
    \label{Fig:Nt4SparsePolyakovBinder}
\end{figure}

\section{Autocorrelation and ensemble information cards}
\label{App:Autocorrelation}

The generation of ensembles using a Markov chain proceeds via local updates of the link variables that produce correlated trajectories. In order to have statistically significant ensembles of non correlated configurations, we must measure the autocorrelation, $\tint$,  among  trajectories. For a pedagogical reference we refer the reader to Ref.~\cite{Ambegaokar:2010a}. Then, one can either select a subset of the configurations to retain by sampling them so that their separation in simulation time is much larger than the autocorrelation, or, equivalently, by binning them in bin sizes larger than the autocorrelation time. For this study, we group the trajectories into bins of size $2\tint$, and apply the bootstrap method.

We define the normalised autocorrelation function, for a given observable, $O$, and at a time lag $\tau$ as
\begin{equation}
    C_O(\tau) \equiv 
    \frac{1}{N - \tau}
    \sum_{i=1}^{N-\tau}
    \frac{(O_i -\langle O\rangle)(O_{i+\tau} - \langle O\rangle)}{\mathrm{var}(O)}\,,
    \label{eq:AutocorrelationFunction}
\end{equation}
where $\mathrm{var}(O)$ is the variance of the observable, $\langle O\rangle$ is the expectation value of the operator $O$, and $O_i$ is the $i$-th measurement of the operator. We then
define the integrated autocorrelation time:
\begin{equation}
    \tint \equiv
    \frac{1}{2} + \sum_{\tau=1}^\infty C(\tau)\,.
    \label{eq:IntergratedAutocorrelationTime}
\end{equation}
Note the summation over the different time lags, $\tau$, ranging to $\infty$. Samples may remain correlated even at large time lags, and the integrated autocorrelation time describes the correlation structure of an ideal, infinitely long Markov chain. In practice, the number of measurements in a chain is finite, and we must find a way to truncate the number of terms in the sum.

For a finite sample, the sum in Eq.~(\ref{eq:IntergratedAutocorrelationTime}) is truncated to a  window of finite length, $W$.
We use the Madras-Sokal self-consistent windowing procedure~\cite{Madras:1988ei}, to set the window size. This is implemented as follows: the size of the window is set by the relation
\begin{equation}
    W \equiv  \lfloor c \cdot\tint \rfloor\,,
    \label{eq:MadrasSokalWindowLength}
\end{equation}
with $\lfloor x \rfloor$ being the integer floor value of a real number $x$ and where we have chosen $c=5$. Starting from an initial guess for the integrated autocorrelation time, $\tint^0 = 0.5$,  and iteratively computing $\tint$ using the corresponding window size $W^{(k)}$ from~\eqref{eq:MadrasSokalWindowLength}:
\begin{equation}
    \tint^{(k+1)} \equiv 
    \frac{1}{2} + \sum_{\tau=1}^{W^{(k)}} C(\tau)\,,
    \label{eq:MadrasSokalWindow}
\end{equation}
until the following convergence criterion is satisfied
\begin{equation}
    | \tint^{(k+1)} - \tint^{(k)} | \le \epsilon\,,
    \label{eq:MadrasSokalConvergenceCriterion}
\end{equation}
where $\epsilon$ is a chosen tolerance, which we have conventionally set to $\epsilon=10^{-5}$.

In the vicinity of phase transitions, the challenges arising in treating non-ergodicity phenomena in the Markov chain are exacerbated by the fact that the probability of tunnelling between different vacua is suppressed by the volume---see for example the lattice estimates of the critical point of the ${\rm SU}(3)$ theory with two fundamental fermions, in Ref.~\cite{Cuteri:2020yke} and references therein.

Given the large number ensembles analysed for this study,  we provide details about them in the data release~\cite{data_release}. We find it useful to show in Fig.~\ref{Fig:EnsembleCards} two representative ensemble information cards, both pertaining ensembles with $am_0=4$ and $N_t=4$, and with a representative value of the coupling, $\beta=7.3375$, chosen to lie in close proximity to the transition. The two cards differ by the choice of spatial extent of the lattice, allowing to compare the smallest, $N_s=8$, and largest, $N_s=24$, available.  

The cards show a number of observables and their evolution along the trajectories: the history of the plaquette, $P$, and Polyakov loop, $\Phi$, are shown next to their autocorrelation length, and their histogram. 

We are particularly interested in monitoring the autocorrelation time of our ensembles. For this reason we compare our estimate of the integrated autocorrelation time with the autocorrelation time calculated via the binning method. In the latter, the measurements are binned together in bins of increasing size, measuring the estimated statistical error on the sample as a function of the bin size. This value is expected to saturate and to reach a plateau when the bin size is of the order of the autocorrelation time. 

We implement this consistency check on our estimated value of $\tint$, in the bottom pane of each ensemble card, where we show the ratio $\frac{\Delta |\Phi|}{\Delta |\Phi^0|}$. The numerator $\Delta |\Phi|$ is the standard deviation of the absolute value of the Polyakov loop with the measurements binned at increasing bin size (on the horizontal axis). The denominator $\Delta |\Phi^0|$ is the standard deviation of the unbinned measurements of the Polyakov loop absolute value.

Individual cards of this type are generated and inspected for all ensembles used in our study, as a way to gain insight into the level of ergodicity of the algorithm, and to ascertain the absence of glaring anomalies, which would otherwise signal the presence of possible systematics or failures in the algorithm. We observe the presence of the triple-peak histogram structure, typical of  first-order phase transitions involving the breaking of an approximate $\mathbb{Z}_2$ symmetry, for $V=(8a)^3$. We also notice that tunnelling becomes increasingly difficult at larger lattice volumes, so that this feature is less prominent, and requires larger ensembles to be resolved numerically.

\section{Binder cumulant}
\label{Sec:Binder}

The Binder cumulant is defined as follows:
\begin{equation}
    B_4(X) \equiv 1 - \frac{\langle X^4\rangle}{3\langle X^2 \rangle ^2}\,,
    \label{eq:BinderCumulant}
\end{equation}
where $X$ is an observable (thermodynamic quantity) of interest, which in our case is the Polyakov loop, $\Phi$---or, equivalently, its absolute value, $|\Phi|$, as the observable enters only quadratically in Eq.~(\ref{eq:BinderCumulant}). This quantity provides an established way of estimate the critical coupling, $\beta_c$. For values of $\beta$ far from the phase transition (and in the limit of  large volume), the Binder cumulant, $B_4(\Phi)$, is predicted to be $B_4(\Phi)=2/3$. A suppression of this value is expected to emerge in the vicinity of the critical coupling, $\beta_c$,  that hence can be estimated by measuring the position of the minimum. This determination is expected to agree with other determinations of the critical coupling in the thermodynamic limit, $V\rightarrow +\infty$.

We report in Fig.~\ref{Fig:Nt4SparsePolyakovBinder} the Binder cumulant of the Polyakov loop, for the five sets of ensembles discussed in the main body of the paper, in the case of lattices with the smallest available temporal extent, $N_t=4$, and using multiple histogram reweighting. The results clearly show a qualitative difference between the case of first-order phase transitions and those of smooth crossovers. Due to the statistical and systematic uncertainties of these measurements, the determination of $B_4(\Phi)$ is rather noisy, yet the position of the critical coupling is compatible with high precision Yang-Mills measurements taken from  the literature. For reference, the estimate of critical coupling, in the Yang-Mills ${\rm Sp}(4)$ theory, has been computed for the plaquette, $X= P$, with $N_s=28$,  for $N_t=4$, to yield $\beta_c(B_4)\simeq 7.340024(17)$~\cite{Bennett:2025whm}.
The measurement of the Binder cumulant of  the Polyakov loop, $X=\Phi$, can be found in Ref.~\cite{Bruno:2024dha}, and  for $N_t=4$, it yields $\beta_c(B_4)\simeq 7.339(1)$.

\bibliographystyle{JHEP}
\bibliography{ref}

\end{document}